\documentclass[12pt]{article}

\makeatletter
\@addtoreset{equation}{section}

\makeatother

\begin{document}

\title{Algebraic algorithm for the computation of one-loop Feynman 
diagrams in lattice QCD with Wilson fermions}
\author{
  \\
  {\small Giuseppe Burgio}              \\[-0.2cm]
  {\small\it Dipartimento di Fisica and INFN -- Sezione di Parma}  \\[-0.2cm]
  {\small\it Universit\`a degli Studi di Parma}        \\[-0.2cm]
  {\small\it I-43100 Parma, ITALIA}          \\[-0.2cm]
  {\small Internet: {\tt BURGIO@PARMA.INFN.IT}}     \\[-0.2cm]
  \\[-0.1cm]  \and
  {\small Sergio Caracciolo}              \\[-0.2cm]
  {\small\it Dipartimento di Fisica and INFN -- Sezione di Lecce}  \\[-0.2cm]
  {\small\it Universit\`a degli Studi di Lecce}        \\[-0.2cm]
  {\small\it I-73100 Lecce, ITALIA}          \\[-0.2cm]
  {\small Internet: {\tt CARACCIO@UX1SNS.SNS.IT}}     \\[-0.2cm]
  \\[-0.1cm]  \and
  {\small Andrea Pelissetto}                          \\[-0.2cm]
  {\small\it Dipartimento di Fisica and INFN -- Sezione di Pisa}    \\[-0.2cm]
  {\small\it Universit\`a degli Studi di Pisa}        \\[-0.2cm]
  {\small\it I-56100 Pisa , ITALIA}          \\[-0.2cm]
  {\small Internet: {\tt PELISSET@IBMTH1.DIFI.UNIPI.IT}}   \\[-0.2cm]
  {\small \hphantom{Internet: }{\tt 
           PELISSET@IPIFIDPT.DIFI.UNIPI.IT}}   \\[-0.2cm]
  {\protect\makebox[5in]{\quad}}  
  \\
}
\vspace{0.5cm}

\maketitle
\thispagestyle{empty}   

\vspace{0.2cm}

\begin{abstract}
We describe an algebraic algorithm which allows to express every one-loop
lattice integral with gluon or Wilson-fermion propagators in terms of 
a small number of basic constants which can be computed with 
arbitrary high precision. Although the presentation is restricted to
four dimensions the technique can be generalized to 
every space dimension. Various examples are given, including 
the one-loop self-energies of the quarks and gluons and the 
renormalization constants for some dimension-three and dimension-four 
lattice operators. We also give a method to express the lattice
free propagator for Wilson fermions in coordinate space as a linear function
of its values in eight points near the origin. This is an essential step
in order to apply the recent methods of L\" uscher and Weisz to 
higher-loop integrals with fermions.
\vskip 0.1truecm

\end{abstract}

\clearpage

\def\smfrac#1#2{{\textstyle\frac{#1}{#2}}}
\newcommand{\be}{\begin{equation}}
\newcommand{\ee}{\end{equation}}
\newcommand{\<}{\langle}
\renewcommand{\>}{\rangle}

\def\spose#1{\hbox to 0pt{#1\hss}}
\def\ltapprox{\mathrel{\spose{\lower 3pt\hbox{$\mathchar"218$}}
 \raise 2.0pt\hbox{$\mathchar"13C$}}}
\def\gtapprox{\mathrel{\spose{\lower 3pt\hbox{$\mathchar"218$}}
 \raise 2.0pt\hbox{$\mathchar"13E$}}}

\def\bsigma{\mbox{\protect\boldmath $\sigma$}}
\def\bpi{\mbox{\protect\boldmath $\pi$}}
\def\btau{\mbox{\protect\boldmath $\tau$}}
\def\hatp{\hat p}
\def\hatl{\hat l}

\def\msbar{ {\overline{\hbox{\scriptsize MS}}} }
\def\normalmsbar{ {\overline{\hbox{\normalsize MS}}} }

\newcommand{\R}{\hbox{{\rm I}\kern-.2em\hbox{\rm R}}}

\newcommand{\reff}[1]{(\ref{#1})}

\section{Introduction}

Perturbation theory plays an important role in our present understanding of
quantum field theory. In particular on the lattice Feynman diagram computations
are performed to obtain such quantities as ratios
of $\Lambda$-parameters, non-universal coefficients of $\beta$-functions
or renormalization constants of lattice operators. Due to the loss of 
Lorentz invariance lattice calculations are usually particularly involved
and thus in order to obtain reliable results one has to make use of 
computer symbolic programs. To implement this strategy an important step
is the simplification of the lattice integrals appearing in the 
Feynman diagrams of the theory. In \cite{CMP} we presented a general 
technique which allows to express every one-loop bosonic integral 
at zero external momentum 
in terms of two unknown basic quantities which could be computed
numerically with high precision. This method allows the complete 
evaluation of every diagram with gluon propagators. 

In this paper we want to generalize the technique to deal with integrals
with both gluonic and fermionic propagators. For the fermions we use the 
Wilson action \cite{Wilson}. 
Notice that our method depends only on
the structure of the propagator and thus it can be applied in calculations
with the standard Wilson action as well as with the improved
clover action \cite{clover-action}. 
We show that every integral at zero external momentum 
can be expressed in terms of 
a small number of basic quantities (nine for purely fermionic integrals,
thirteen for integrals with bosonic and fermionic propagators). The 
advantage of this procedure is twofold: first of all every Feynman
diagram can be computed in a completely symbolic way making it easier
to perform checks and verify cancellations; moreover the basic 
constants can be easily computed with high precision and thus the 
numerical error on the final result can be reduced at will.
Although the presentation is restricted to
four dimensions the technique can be generalized to
every space dimension.

A second important application of our method is connected with the use 
of coordinate-space methods for the evaluation of higher-loop 
Feynman diagrams. This technique, introduced by 
L\"uscher and Weisz \cite{LW}, is extremely powerful and allows  a
very precise determination of two- and higher-loop integrals.
One of the basic ingredients of this method is the computation of the 
free propagator in coordinate space. We will give an algorithm
which allows the analytic determination of the free fermionic propagator in 
$x$-space in terms of eight basic constants which can be reinterpreted as the 
values of the propagator near the origin.

The paper is organized as follows: 
in section 2 we review the computation of the continuum limit
of lattice integrals, showing that the calculation 
can be split in two parts: the evaluation of a subtracted continuum
integral and of a certain number of lattice integrals with zero 
external momentum which are then discussed in the following sections. 
In section 3 we present the method in the 
bosonic case, simplifying the strategy discussed in \cite{CMP}. 
Section 4 represents the core of the paper and gives the detailed 
algorithm for fermionic and mixed bosonic-fermionic integrals. In 
both these sections we need to introduce an infrared cut-off to 
regularize the integrals for $k=0$. We choose to introduce a mass $m$; 
other possibilities are discussed in section 5 where the connection
with dimensionally regularized integrals is presented. 
Finally in section 6 we present a few examples: we give the analytic
expressions for the lattice gluon and fermion self-energy, 
for the renormalization constants of dimension-three bilinear fermion
operators and a computation of the renormalization constants for the 
operators which show up in the energy-momentum tensor and which 
are relevant in the computation of the structure functions 
which appear in
the deep-inelastic scattering \cite{Buras}. Finally we discuss 
an algorithm for the computation of the free lattice propagators.

\section{Continuum limit of lattice integrals}

\newcommand{\B}{{\cal B}}
In this Section we want to discuss the computation of one-loop
Feynman diagrams on the lattice. In general each
graph has the form
\be
G(\{ p_i\}) = \int {d^4 q\over (2 \pi)^4} F(q;\{ p_i\})
\label{eq2.1}
\ee
where $\{ p_i\}$ is a set of external momenta. Of course one is not
interested in the exact computation of \reff{eq2.1} but only in its
value in the continuum limit. If the integral is ultraviolet-convergent
one can simply substitute $F(q;\{p_i\})$ with its continuum counterpart
and obtain a continuum convergent integral. 
Let us now suppose that \reff{eq2.1} is divergent and, for simplicity,
that there is only one external momentum. 
If every propagator
is massive so that $F(q;\{ p_i\})$ is finite for any set of momenta
going to zero, one can use the general technique of 
BPHZ \cite{Collins} which have been generalized on the lattice 
by Reisz \cite{Reisz} writing 
\begin{eqnarray}
G(p) &=& \int {d^4 q\over (2 \pi)^4}  
   \left[ F(q;p) - (T^{n_F} F)(q;p)\right] + 
     \int {d^4 q\over (2 \pi)^4} (T^{n_F} F)(q;p) \nonumber \\
  &\equiv & G^c(p) +\, G^L (p) 
\label{eq2.2}
\end{eqnarray}
where $n_F$ is the degree of the divergence of the integral and 
\be
   (T^{n_F} F)(q;p) = \sum_{k=0}^{n_F} 
     {1\over k!} p_{\mu_1} \ldots p_{\mu_k} 
     \left[ {\partial\over \partial p_{\mu_1} } \ldots
            {\partial\over \partial p_{\mu_k}} F(q;p)
             \right]_{p=0}
\ee
The first integral in \reff{eq2.2} is ultraviolet-finite \cite{Reisz} 
and thus one can take the 
continuum limit. Thus all the effects of the lattice regularization
remain only in the second term, which is simply a polynomial
in the external momentum with coefficients given by lattice
zero-momentum integrals. 

If the integrand contains massless propagators
one has to be more careful: indeed an expansion around $p=0$ 
can give rise to infrared divergences. A simple way out consists
in introducing an intermediate infrared regularization: 
one can use the 
dimensional regularization \cite{Kawai,CMP} working in 
dimension $d>4$, or introduce a mass in the propagators. In both
cases $G^c(p)$ and $G^L(p)$ will be singular for $d\to 4$
or $m\to 0$ but, of course, the singularity will cancel 
when summing up the two terms.

In conclusion the computation of the continuum limit of \reff{eq2.1}
splits in the computation of two different quantities: 
a continuum ultraviolet-finite integral and a certain number 
of zero-momentum lattice integrals whose computation will be 
discussed in the next sections.

\section{Bosonic integrals}

In this section we discuss the evaluation of the most general one-loop
lattice integral at zero external momentum with bosonic 
propagators. It is very easy to see that any such integral 
can be written as a linear combination of terms of the form
\be
\B(p;n_x,n_y,n_z,n_t) = 
\int_{-\pi}^\pi {d^4 k\over (2 \pi)^4}  \, {
\hat{k}^{2 n_x}_x \hat{k}^{2 n_y}_y \hat{k}^{2 n_z}_z \hat{k}^{2 n_t}_t\over
D_B(k,m)^p } \label{Bint}\ee
where $p$ and $n_i$ are positive integers, $\hat{k}_\mu = 2 \sin(k_\mu/2)$
and 
\be
D_B(k,m) = \hat{k}^2 + m^2 \; .
\label{bosonicprop}
\ee
is the inverse bosonic propagator.
In the following when one of the arguments $n_i$ is zero it will be 
omitted as argument of $\B$. 

We will review here a general technique for expressing these integrals 
in terms of three constants \cite{CMP}.

\newcommand\BB{\B_\delta}

We will first generalize (\ref{Bint}) by considering the following 
more general integrals
\be
\BB(p;n_x,n_y,n_z,n_t) = 
\int_{-\pi}^\pi {d^4k\over (2 \pi)^4}  \, {
\hat{k}^{2 n_x}_x \hat{k}^{2 n_y}_y  \hat{k}^{2 n_z}_z \hat{k}^{2 n_t}_t\over
D_B(k,m)^{p+\delta} } 
\ee
where $p$ is an arbitrary integer (not necessarily positive) 
and $\delta$ a real number which
is introduced in order to avoid singular cases at intermediate stages of the 
computation and which will be set to zero at the end. 

The first result we want to prove is that each integral 
$\BB(p;n_x,n_y,n_z,n_t)$
can be reduced through purely algebraic manipulations to a sum of integrals 
of the same type with $n_x=n_y=n_z=n_t=0$.

Indeed the integrals $\BB$ satisfy the following recursion relations:
\begin{eqnarray}
\BB(p;1) & = & {1\over 4} \,[\BB(p-1) - m^2 \BB(p)]	\nonumber\\ [1mm]
\BB(p;x,1) & = & {1\over 3}\,[\BB(p-1;x) - \BB(p;x+1) - m^2 \BB(p;x) ] 
        \nonumber \\ [1mm]
\BB(p;x,y,1) & = & {1\over 2}\,[\BB(p-1;x,y) - \BB(p;x+1,y) - 
            \BB(p;x,y+1) \nonumber \\ 
       && \qquad \qquad - m^2 \BB(p;x,y) ] \nonumber \\ [1mm]
\BB(p;x,y,z,1) & = & \BB(p-1;x,y,z) - \BB(p;x+1,y,z) - 
            \BB(p;x,y+1,z) \nonumber \\ 
       && \qquad \qquad - \BB(p;x,y,z+1) - m^2 \BB(p;x,y,z)  
\label{bosonicrec1}
\end{eqnarray}
which can be obtained using the trivial identity
\be
D_B(k,m)\, = \sum_{i=1}^4 \hat{k}^2_i +\, m^2 \;\; .
\ee
Furthermore, when $r>1$, we can write 
\be
{({\hat k}^2_w )^r \over D_B(k,m)^{p+\delta}} = 
4 {({\hat k}^2_w )^{r-1} \over D_B(k,m)^{p+\delta}} +
2 {({\hat k}^2_w )^{r-2} \over p + \delta - 1} \sin k_w 
{\partial \over \partial k_w} { 1 \over D_B(k,m)^{p+\delta-1}} 
\ee
Then, integrating by parts, we obtain the recursion relation:
\begin{eqnarray}
\lefteqn{\BB(p;\ldots,r)} \nonumber \\ 
         & = & {r-1\over p+\delta-1}
	\BB(p-1;\ldots,r-1) -
	{4r-6\over p+\delta-1} \BB(p-1;\ldots,r-2) \nonumber \\
	&	& +\, 4\, \BB(p;\ldots,r-1) \label{rec2}
\end{eqnarray}
Let us notice that for $p\not=1$ this recursion relation is regular for
$\delta\to 0$. For $p=1$ instead the coefficients of $\BB(0;\ldots)$ diverge
as $1/\delta$. This means that to compute $\BB(1;\ldots)$ for $\delta=0$, 
we need to compute $\BB(0;\ldots)$ including terms of order $\delta$. 
Since by the application of the previous recursion relations to 
$\BB(0;\ldots)$ we generate $\BB(-1;\ldots)$, then $\BB(-2;\ldots)$ and 
so on and their coefficients are finite for $\delta\to 0$, we see 
that in general we need to compute all integrals 
$\BB(p;n_x,n_y,n_z,n_t)$ with $p\le 0$ up to $O(\delta^2)$. 

The previous relations allow to reduce every integral $\BB(p;n_x,n_y,n_z,n_t)$
to a sum of the form 
\be
\BB(p;n_x,n_y,n_z,n_t) = \sum_{r=p-n_x-n_y-n_z-n_t}^{p} 
a_r (m,\delta) \BB(r)
\label{BBn}
\ee
The $m$-dependence of $a_r (m,\delta)$ is very simple: it is indeed 
a polynomial in $m^2$. 
Let us now discuss the $\delta$-dependence. If $p\le 0$ only nonpositive
values of $r$ are allowed in \reff{BBn} with coefficients which are 
regular for $\delta\to 0$. For $p>0$ the situation is more complicated:
for $r\ge 1$, $\lim_{\delta\to 0}$ $a_r(m,\delta)$ is finite 
while
for $r\le0$ $a_r(m,\delta)$ may behave as $1/\delta$ 
when $\delta$ goes to zero. As we already observed,
this means that we need to compute $\BB(r)$, $r\le 0$, 
including terms of order $\delta$.

Now let us show that all $\BB(p)$ can be expressed in terms 
of a finite number of them.
Although this can be shown for generic values of the mass\footnote{The 
analogous procedure in two dimensions is presented in Appendix A.3 
of \cite{CP-4loop}.} many simplifications occur if one restricts the 
attention to the massless case, i.e. if one considers the limit
$m^2\to 0$ keeping only the non-vanishing terms.  Let us thus discuss 
the limit $m^2\to0$ of $\BB(p)$.

Simple power-counting shows that the integrals $\BB(p)$, $p\le 1$, are finite.
To compute the divergent part of $\BB(p)$, $p\ge 2$, let us start from 
the well-known representation 
\be
\BB(p)\, =\, {1\over 2^{p+\delta} \Gamma(p+\delta)} 
\int_0^\infty d\lambda \, \lambda^{p+\delta-1} e^{-m^2\lambda/2-4 \lambda}
\, I_0(\lambda)^4
\ee
where $I_0(\lambda)$ is a modified Bessel function \cite{Gradshteyn}.
We introduce
constants $b_i$ that are defined by the asymptotic expansion for 
large $x$ of $I_0(x)^4$ as 
\be
I_0(x)^4 \approx 
e^{4 x} \sum_{i=0}^\infty {b_i\over x^{i+2}} \;\; .
\ee
The constants $b_i$ are rational numbers multiplied by $1/\pi^2$.
Then the divergent part of $\BB(p)$ is given by
\be
{1\over \Gamma(p+\delta)} \sum_{i=2}^p {b_{i-2} \Gamma(p+\delta-i)\over
         2^i (m^2)^{p+\delta-i} }
\ee
Since here $p$ is positive we can set $\delta=0$. However in the next section
we will be interested also in the divergent part proportional to 
$\delta$. Thus expanding in $\delta$ and neglecting finite terms, 
up to $O(\delta^2)$, we get\footnote{Notice that we are dealing here
with two different limits, $m^2\to 0$ and $\delta\to 0$. In general they
do not commute and thus it is necessary to specify the correct order
in which they are taken. Here we first  consider $\delta\to 0$ at fixed
$m$ and then we let $m$ go to zero.}
\begin{eqnarray}
&& {1\over \Gamma(p)} 
   \sum_{i=2}^{p-1} {b_{i-2} \Gamma(p-i)\over 2^i (m^2)^{p-i}}
   -\, {b_{p-2}\over 2^p \Gamma(p)} \log m^2 \nonumber \\
&& - \delta\left[ {1\over \Gamma(p)} \sum_{i=2}^{p-1} 
      {b_{i-2} \Gamma(p-i)\over 2^i (m^2)^{p-i}} 
      \left( \log m^2 + \sum_{k=p-i}^{p-1} {1\over k}\right)\right. 
      \nonumber \\
&& \qquad \left. - \, {b_{p-2}\over 2^p \Gamma(p)} 
    \left({1\over2} \log m^2 + \sum_{k=1}^{p-1} {1\over k}\right) \log m^2
   \right] +\, O(\delta^2)
\label{divergenzabosonico}
\end{eqnarray}
Let us now go back to \reff{BBn} rewriting it as 
\be
\BB(p;n_x,n_y,n_z,n_t) \, =\, 
   \sum_r a_r(0,\delta) \BB(r) + 
   \sum_r (a_r(m,\delta) - a_r(0,\delta)) \BB(r)
\label{eq2.13}
\ee
and let us consider the limit $m^2\to 0$. It is clear that in the second
sum only $\BB(r)$ with $r\ge 3$ can contribute, since only these integrals
have power divergences for $m^2\to 0$. As the values of $r$ in the sums 
satisfy $r\ge p$, we find that the second sum contributes only for 
$p\ge 3$. Thus we get for $p\le 2$
\be
\BB(p;n_x,n_y,n_z,n_t) \, =\,
   \sum_r a_r(0,\delta) \BB(r) 
  +\, O(m^2)
\label{BBnOm2uno}
\ee
while, for $p > 2$,
\be
\BB(p;n_x,n_y,n_z,n_t) \, =\,
   \sum_r a_r(0,\delta) \BB(r) +\, {\cal R}(m,\delta)
  +\, O(m^2)
\label{BBnOm2due}
\ee
where ${\cal R}(m;\delta)$ is 
a polynomial in $1/m^2$ whose coefficients, for $\delta\to 0$,
are rational numbers multiplied by $1/\pi^2$.

Let us finally find the last recursion relations. Let us start from
the trivial identity
\be
\BB(p;1,1,1,1)\, -\, 4\BB(p+1;2,1,1,1) \, -m^2\,\BB(p+1;1,1,1,1)\, =\, 0
\label{identitabosonica}
\ee
and let us apply the previous procedure to reduce each term to the 
form \reff{BBnOm2uno} and \reff{BBnOm2due}\footnote{Since 
we need to compute $\BB(r)$, $r\le 0$,
including terms of order $\delta$, in the computation of the various terms,
one must keep the contributions of order $\delta$ if $p<0$, while for 
$p\ge 0$ it is enough to expand the identity to order $O(\delta)$.}. 
We thus get a non trivial relation involving
$\BB$ of the form
\be
\sum_{r=p-4}^p b_r(p;\delta) \BB(r) +\, {\cal S}(p;m,\delta) = 0
\label{idbos}
\ee
where ${\cal S}(p;m,\delta)=0$ for $p\le 2$ while for $p>2$ is 
a polynomial in $1/m^2$ which is finite for
$\delta\to 0$. 
We will use this identity to express all $\BB(p)$ in terms
of $\BB(r)$, $0\le r\le 3$. Indeed we can solve \reff{idbos} in terms
of $\BB(p)$ and thus we get a relation which expresses it in terms
of $\BB(p-1),\ldots,\BB(p-4)$. We will use this relation for 
$p \ge 4$. On the other hand we can solve \reff{idbos} in terms 
of $\BB(p-4)$ and then shift $p\to p+4$. In this way we obtain a relation
which expresses $\BB(p)$ in terms of $\BB(p+1),\ldots,\BB(p+4)$. We
use this recursion for $p\le -1$. Applying recursively these two
relations we get finally  ($p\not=0,1,2,3$)
\be
\BB(p) \, =\, \sum_{r=0}^3 c_r(p;\delta) \BB(r) +\, {\cal T}(p;m,\delta)
\ee
where ${\cal T}(p;m,\delta)$ is a polynomial in $1/m^2$.
A direct analysis of \reff{idbos} shows the following 
properties:
\begin{enumerate}
\item if $p\ge 4$, $c_0(p;\delta) = O(\delta^4)$;
\item if $p \le -1$, $c_r(p;\delta) = O(\delta)$ for $1\le r \le 3$;
\item if $p \le -1$, ${\cal T}(p;m,\delta)$ is of order $\delta$, while
for $p\ge 4$ it is finite for $\delta\to 0$.
\end{enumerate}
Substituting back in \reff{BBnOm2uno} or \reff{BBnOm2due} we get
\be
\BB(p;n_x,n_y,n_z,n_t) \, =\,
A(\delta) \BB(0) + B(\delta) \BB(1) + 
C(\delta) \BB(2) + D(\delta) \BB(3) + E(m,\delta) 
\ee
where $E(m,\delta)$ is a polynomial in $1/m^2$. 

We can now go back to the original integral \reff{Bint} (notice that
we are only interested in the case $p>0$). 
Because of the second property of the coefficients $c_r(p;\delta)$
and the property of ${\cal T}(p;m,\delta)$, 
we immediately see that $B(\delta)$, $C(\delta)$, $D(\delta)$ 
and $E(m,\delta)$ are 
finite for $\delta\to 0$. Then, as the l.h.s. is obviously finite 
for $\delta\to 0$, also $A(\delta)$ is finite for 
$\delta\to 0$. Since ${\cal B}(0) = 1$, we have finally
\be
{\cal B}(p;n_x,n_y,n_z,n_t) \, =\,
A(0) + B(0) {\cal B}(1) + 
C(0) {\cal B}(2) + D(0) {\cal B}(3) + E(m,0) 
\ee
We want now to make contact with our previous work where all
results were expressed in terms of three constants,
$Z_0$, $Z_1$ and $F_0$ defined by
\begin{eqnarray}
Z_0 &=& \left. {\cal B}(1) \right|_{m = 0} \\
Z_1 &=& {1\over 4} \left. {\cal B}(1;1,1) \right|_{m = 0} \\
F_0 &=& \lim_{m\to 0} [16\pi^2 {\cal B}(2) + \log m^2 + \gamma_E ]
\end{eqnarray}
It is clear how to rewrite ${\cal B}(1)$ and ${\cal B}(2)$ in terms of 
$F_0$ and $Z_0$. For ${\cal B}(3)$ a short calculation gives 
\be
{\cal B}(3) =\, {1\over 32 \pi^2 m^2} - {1\over 128 \pi^2} 
    \left(\log m^2 + \gamma_E - F_0\right) - {1\over 1024}
    - {13 \over 1536 \pi^2} + {Z_1\over 256}
\ee
An additional simplification occurs if the original integral is finite.
In this case the $\log m^2$ terms must cancel. They appear only in 
${\cal B}(2)$ and ${\cal B}(3)$ and always in the combination 
$(\log m^2 + \gamma_E - F_0)$. Thus the cancellation of $\log m^2$ 
implies also the cancellation of $\gamma_E$ and $F_0$. All finite integrals
are thus functions of $Z_0$ and $Z_1$ only. Numerical values of 
the constants are reported in Table \ref{costanti_bosoniche}.
\begin{table}
\begin{center}
\begin{tabular}{|c|l|}
\hline
$Z_0$ & 0.154933390231060214084837208 \\
$Z_1$ & 0.107781313539874001343391550 \\
$F_0$ & 4.369225233874758 \\
\hline
\end{tabular}
\end{center}
\caption{Numerical values of the three constants $Z_0$, $Z_1$ and $F_0$.
}
\label{costanti_bosoniche}
\end{table}

It is interesting to notice that the same technique can be used for bosonic 
integrals in $d$ dimensions. The basic recursions can be trivially generalized
as well as the identity \reff{identitabosonica}. For generic $d$ 
we find that all integrals can finally be expressed in terms of 
$\BB(1)$, $\ldots$, $\BB(d-1)$, i.e. in terms of $(d-1)$ constants,
reducing to $(d-2)$ for infrared finite integrals.

\newcommand\FF{{\cal F}_\delta}
\section{Integrals with bosonic and Wilson-fermion propagators}

We want now to discuss the computation of general integrals with
fermionic and bosonic propagators at zero external momenta. Define
\be
{\cal F} (p,q;n_x,n_y,n_z,n_t) =\,
 \int_{-\pi}^\pi {d^4 k \over (2\pi)^4}
 {\hat{k}_x^{2 n_x} \hat{k}_y^{2 n_y} \hat{k}_z^{2 n_z} \hat{k}_t^{2 n_t}\over
     D_F(k,m_f)^p D_B(k,m_b)^q}
\ee
where $p$, $q$ and $n_i$ are positive integers,
$D_B(k,m_b)$ is defined in \reff{bosonicprop} and
\be
D_F(k,m_f) = \sum_i \sin^2 k_i + {r^2_W\over 4} (\hat{k}^2)^2 + m_f^2
\label{Fermiprop}
\ee
is the denominator appearing in the 
propagator for Wilson fermions\footnote{To be precise, 
$D_F(k,m_f)$ is the denominator
in the propagator for Wilson fermions only for $m_f=0$.
For $m_f \not =0 $ the correct denominator would be
\be
\hat{D}_F(k,m_f) = \sum_i \sin^2 k_i + \left({r_W\over 2} \hat{k}^2 + m_f
    \right)^2 \;\; .
\label{trueFermiprop}
\ee
However in our discussion $m_f$ will only play the role of an infrared
regulator and thus it does not need to be the {\em true} fermion mass.
The definition \reff{Fermiprop} is easier to handle than
\reff{trueFermiprop}. For a discussion of integrals using 
\reff{trueFermiprop} see Section 5.2.}.
In the following when one of the arguments $n_i$ is zero it will be
omitted as an argument of $\cal F$.
Following the strategy we have used in the purely bosonic case we will
first generalize the integrals introducing
\be
\FF (p,q;n_x,n_y,n_z,n_t) =\,
 \int_{-\pi}^\pi {d^4 k\over (2\pi)^4}
 {\hat{k}_x^{2 n_x} \hat{k}_y^{2 n_y} \hat{k}_z^{2 n_z} \hat{k}_t^{2 n_t}\over
     D_F(k,m_f)^{p+\delta} D_B(k,m_b)^q}
\ee
Here we consider both $p$ and $q$ as arbitrary integers (not necessarily
positive).  The parameter $\delta$ 
is used in the intermediate steps of the calculation and will be set to zero
at the end.

To simplify the discussion we will only consider the case $r_W=1$
but the technique can be applied to every value of $r_W$. Moreover
we will restrict our attention to the massless case, i.e. we will 
consider the integrals $\FF$ in the limit $m_b=m_f\equiv m\to 0$.

In the following we will present a procedure that allows to compute iteratively
a generic $\FF$ in terms of a finite number of them: precisely 
every $\FF(p,q;n_x,n_y,n_z,n_t)$ with $q\le 0$ can be expressed
in terms of $\FF(1,0)$, $\FF(1,-1)$, $\FF(1,-2)$, $\FF(2,0)$,
$\FF(2,-1)$, $\FF(2,-2)$, $\FF(3,-2)$, $\FF(3,-3)$ and $\FF(3,-4)$;
the integral $\FF(2,0)$ appears only in infrared-divergent integrals.
If $q>0$ the result contains three additional constants together with 
the bosonic quantities $Z_0$, $Z_1$ and $F_0-\gamma_E$.

Our procedure works in four steps:
\begin{enumerate}
\item First step: we express each integral $\FF(p,q;n_x,n_y,n_z,n_t)$ in terms
of $\FF(p,q)$ only.
\item Second step: we express every $\FF(p,q)$ in terms of 
$\FF(r,s)$ with $0\le r \le 3$, arbitrary $s$ or $r \leq -1$ and 
$s=1,2,3$ or $r\ge 4$ and $s=0,-1,-2$. This is obtained by a systematic use 
of the identity
\be
 {\cal I}_1 (p,q) \equiv
   \FF (p,q;1,1,1,1) - 4 \FF (p,q+1;2,1,1,1) -
   m^2 \FF (p,q+1;1,1,1,1) =\, 0    
\label{identity1}
\ee
\item Third step: we express the remaining $\FF(p,q)$ in terms of 
$\FF(r,s)$ with $r=3$, $-4\le s \le 0$ or $r=2$, $-4\le s \le 2$
or $r=1$, $-4\le s \le 4$ or $r=0$, $-4\le s \le 6$ or $r=-1$ and $s=2$.
This is obtained by a systematic use of the identity 
\begin{eqnarray}
&& \hskip -40pt {\cal I}_2 (p,q) \equiv
   \FF (p,q;1,1,1,1) - \FF (p+1,q-1;1,1,1,1) +
   \FF (p+1,q;3,1,1,1) - \nonumber \\
&& \qquad {1\over4} \left[
   \FF (p+1,q-2;1,1,1,1) -
   2 m^2 \FF (p+1,q-1;1,1,1,1) \right. \nonumber \\ 
&& \qquad \qquad \left. + m^4 \FF (p+1,q;1,1,1,1)\right] = 0
\label{identity2}
\end{eqnarray}
\item Fourth step: the identities ${\cal I}_1 (p,q)$ and 
${\cal I}_2 (p,q)$ are used to provide additional relations between the 
remaining integrals. We end up with the result we have quoted above.
\end{enumerate}
We want to notice here two basic facts. First of all, as in the bosonic case, 
the structure of the recursion relations, will force us to compute 
all $\FF(p,q;n_x,n_y,n_z,n_t)$ including terms of order $\delta$
when $p\le 0$. As a consequence we will require ${\cal I}_1(p,q)$ and 
${\cal I}_2(p,q)$ to be satisfied up to $O(\delta^2)$ respectively
for $p\le 0$ and $p\le -1$ and up to $O(\delta)$ in the opposite case,
i.e. for $p\ge 1$ and $p\ge 0$ respectively.

Let us finally mention some general properties of all 
the recursion relations we will 
consider: in all cases we  will give results of the form
\be
\FF(p,q;\ldots) \,=\, \sum_{rs} \alpha_{pq;rs} (\delta) \FF(r,s) + 
   {\cal P}_{pq}(m,\delta) + O(m^2)
\ee
where $\alpha_{pq;rs}(\delta)$ and 
${\cal P}_{pq}(m,\delta)$ will always have 
the following properties:
\begin{enumerate}
\item if $q\le 0$, then $\alpha_{pq;rs}(\delta)=0$ for $s>0$. In other words
we will express {\em purely fermionic} integrals in terms of integrals
of the same type;
\item for $\delta\to0$, we have $\alpha_{pq;rs}(\delta)\sim O(1/\delta)$ 
for $p>0$ and $r\le 0$, $\alpha_{pq;rs}(\delta)\sim O(1)$ for 
$p>0$, $r>0$ and $p\le 0$, $r\le 0$ and 
$\alpha_{pq;rs}(\delta)\sim O(\delta)$
for $p\le 0$ and $r>0$;
\item for $\delta\to 0$ we have 
${\cal P}_{pq}(m,\delta) = {\cal P}^{(1)}_{pq}(m)
+ O(\delta)$ for $p>0$ and $q\le 0$, 
${\cal P}_{pq}(m,\delta) = 
\delta {\cal P}^{(1)}_{pq}(m) + O(\delta^2)$ for 
$p\le 0$ and $q\le 0$; for $p>0$ and $q>0$ we have
\be
{\cal P}_{pq}(m,\delta) = 
{1\over\delta}(1 - \delta \log m^2) {\cal P}^{(1)}_{pq}(m) + 
        {\cal P}^{(2)}_{pq}(m) + O(\delta)
\ee
while for $p\le 0$ and $q>0$ we have 
\be
{\cal P}_{pq}(m,\delta) = 
(1 - \delta \log m^2) {\cal P}^{(1)}_{pq}(m) +
       \delta {\cal P}^{(2)}_{pq}(m) + O(\delta^2) \;\; ;
\ee
in all cases ${\cal P}^{(1)}_{pq}(m)$ 
and ${\cal P}^{(2)}_{pq}(m)$ are polynomials
in $1/m^2$ whose coefficients are rational numbers multiplied by $1/\pi^2$.
\end{enumerate}

\subsection{First step: the basic identities}

We will now show that in a purely algebraic way all integrals
can be reduced to a sum of $\FF (p;q)$.
Indeed it is easy to see that these integrals satisfy the following
recursion relations (in giving these recursions we keep $m_b\not=m_f$
and $r_W$ generic)
\begin{eqnarray}
\FF (p,q;1) &=& {1\over4} \left[ \FF (p,q-1) -
     m_b^2 \FF (p,q)\right] \nonumber \\
\FF (p,q;x,1) &=& {1\over3} \left[ \FF (p,q-1;x) -
     m_b^2 \FF (p,q;x) - \FF (p,q;x+1) \right] \nonumber \\
\FF (p,q;x,y,1) &=& {1\over2} \left[ \FF (p,q-1;x,y) -
     m_b^2 \FF (p,q;x,y)\right. \nonumber \\
     && \quad \left. - \FF (p,q;x+1,y) -
         \FF (p,q;x,y+1) \right] \nonumber \\  [2mm]
\FF (p,q;x,y,z,1) &=&  \FF (p,q-1;x,y,z) -
     m_b^2 \FF (p,q;x,y,z) -
     \FF (p,q;x+1,y,z) \nonumber \\
     && \quad - \FF (p,q;x,y+1,z) -
                \FF (p,q;x,y,z+1)
\label{fermionicrec1}
\end{eqnarray}
which can be obtained from the trivial identity
\be
     D_B(k,m_b) = \sum_i \hat{k}_i^2 + m_b^2
\ee
A second recursion relation is obtained from the identity
\be
\sum_i \hat{k}^4_i = 4 (D_B (k,m_b) - D_F(k,m_f) - m_b^2 + m_f^2) +
     r^2_W (D_B(k,m_b) - m_b^2)^2
\ee
In this way we get
\begin{eqnarray}
\FF (p,q;2) &=& \FF(p,q-1) - \FF(p-1,q)
  + (m_f^2 - m_b^2) \FF (p,q) \nonumber \\
     &&\quad + {r^2_W\over4} \left[ \FF (p,q-2) -
		2 m_b^2 \FF (p,q-1) + m_b^4 \FF (p,q)
		\right] \nonumber \\
\FF (p,q;x,2) &=& {4\over3} \left[\vphantom{{1\over4}} \FF (p,q-1;x) -
                 \FF(p-1,q;x) \right. \nonumber \\
     &&\quad \left. + (m_f^2 - m_b^2) \FF (p,q;x) -
            {1\over4} \FF (p,q;x+2) \right] \nonumber \\
     &&\quad + {r^2_W\over3} \left[ \FF (p,q-2;x) -
		2 m_b^2 \FF (p,q-1;x) +
                m_b^4 \FF (p,q;x)
		\right] \nonumber \\ [2mm]
\FF (p,q;x,y,2) &=& 2 \left[\vphantom{{1\over4}}\FF(p,q-1;x,y) -
                                  \FF(p-1,q;x,y) \right. \nonumber \\
      && \quad  + (m_f^2 - m_b^2) \FF (p,q;x,y) -
          {1\over 4} \FF (p,q;x+2,y) \nonumber \\ 
      && \quad \left.  - {1\over 4} \FF (p,q;x,y+2) \right] 
       + {r^2_W\over2} \left[ \FF (p,q-2;x,y) \right. \nonumber \\ 
      && \quad \left. -
		2 m_b^2 \FF (p,q-1;x,y) +
                m_b^4 \FF (p,q;x,y)
		\right] \nonumber \\ [2mm]
\FF (p,q;x,y,z,2) &=& 4 \left[\vphantom{{1\over4}} \FF (p,q-1;x,y,z) -
             \FF(p-1,q;x,y,z) \right. \nonumber \\
        && \quad + (m_f^2 - m_b^2) \FF (p,q;x,y,z) -
               {1\over4} \FF (p,q;x+2,y,z) \nonumber \\
	&& \left. \quad - {1\over4} \FF (p,q;x,y+2,z) -
			  {1\over4} \FF (p,q;x,y;z+2)\right] \nonumber \\
        && \quad + r^2_W \left[ \FF (p,q-2;x,y,z) -
		2 m_b^2 \FF (p,q-1;x,y,z) \right. \nonumber \\
        && \quad \left. +
                  m_b^4 \FF (p,q;x,y,z)
		\right] 
\label{fermionicrec2}
\end{eqnarray}
Finally let us notice that we can write, for $r\ge3$ 
\begin{eqnarray}
{(\hat{k}^2_w)^r\over D_F(k,m_f)^{p+\delta}} & = &
{4 (\hat{k}^2_w)^{r-1} - 4 (2 + r^2_W \hat{k}^2) (\hat{k}^2_w)^{r-3}
     \sin^2 k_w\over
      D_F(k,m_f)^{p+\delta} } \nonumber \\ 
   && \qquad -
{4 (\hat{k}^2_w)^{r-3}\over p + \delta -1} \sin k_w
{\partial\over\partial k_w} {1\over D_F(k,m_f)^{p-1+\delta}}
\end{eqnarray}
Integrating by parts we obtain the recursion relation (to be applied for
$r\ge 3$)
\begin{eqnarray}
&& \FF (p,q;\ldots,r) \nonumber\\ [1mm]
&& \qquad = 6 \FF (p,q;\ldots,r-1) -
            8 \FF (p,q;\ldots,r-2) - \nonumber \\         [2mm]
&& \qquad 4 r^2_W \FF (p,q-1;\ldots,r-2) +
	  4 r^2_W m_b^2 \FF (p,q;\ldots,r-2) + \nonumber\\    [2mm]
&& \qquad r^2_W \FF (p,q-1;\ldots,r-1) -
	  r^2_W m_b^2 \FF (p,q;\ldots,r-1) +  \nonumber\\
&& \qquad {4\over p + \delta - 1} \left[
     -2 q \FF (p-1,q+1,\ldots,r-2) +
     {q\over2} \FF (p-1,q+1;\ldots,r-1) +  \right. \nonumber         \\
&& \qquad \quad \left. (2 r-5) \FF (p-1,q;\dots,r-3) -
     {1\over2} (r-2) \FF (p-1,q;\ldots,r-2)\right]
\label{fermionicrec3}
\end{eqnarray}
Notice that when this recursion is used for $p=1$, terms of the 
form $\FF(0,\ldots)/\delta$ are generated: as we already noticed in the 
introduction to the section, this forces us to compute 
$\FF(p,\ldots)$ including terms of order $\delta$ when $p\le 0$.

Using the previous recursion relations we can now reduce each integral
$\FF (p,q;n_x,n_y,n_z,n_t)$ to a sum of the form
(setting again $m_b=m_f=m$)
\be
\FF (p,q;n_x,n_y,n_z,n_t) \,=\,
  \sum_{r=p-k+1}^p \sum_{s=q-k}^{q+k} a_{rs}(m,\delta) \FF(r,s)
\label{primariduzione}
\ee
where $k= (n_x + n_y + n_z + n_t)$.
It is easy to see, from the structure of the recursion relations, 
that, the coefficients $a_{rs}(m,\delta)$ have the properties mentioned at 
the beginning of the section.

As in the bosonic case we can simplify this expression if we consider the 
limit $m\to 0$.
Let us first compute the divergent 
part of the integrals $\FF(p,q)$ (of course we must have $p+q\ge 2$).
Let us define $\Delta D_F(k,m) \equiv D_F(k,m) - D_B(k,m)$. Then let
us rewrite
\begin{eqnarray}
\FF(p,q) &=& \int {d^4k\over (2\pi)^4} 
    \left[ {1\over D_F(k,m)^{p+\delta} D_B(k,m)^q} \right.  \nonumber \\
    && \qquad  \left. - \, 
           {1\over D_B(k,m)^{p+q+\delta}} 
          \sum_{l=0}^{p+q-2} {-p-\delta\choose{l}}
           \left( {\Delta D_F(k,m)\over D_B(k,m)}\right)^l \right]
       \nonumber \\
     && + \sum_{l=0}^{p+q-2} {-p-\delta\choose{l}} 
        \, \int {d^4k\over (2\pi)^4} 
       { \Delta D_F(k,m)^l\over D_B(k,m)^{p+q+l+\delta} }
\label{divergenzafermionico}
\end{eqnarray}
It is easy to see that the first term is finite. Thus if we want to 
compute the divergences of $\FF(p,q)$ we can limit ourselves to consider the 
second term which contains purely bosonic integrals.
Expanding $\Delta D_F(k,m)^l$ we see that we need to compute 
the divergent part of purely bosonic integrals 
$\BB(r;n_x,n_y,n_z,n_t)$. If the original integral in 
\reff{divergenzafermionico} has $p>0$ we need only the 
$\delta$-independent divergent part, while for $p\le 0$ also the
terms of order $\delta$ are needed. The computation 
of the divergent part of $\BB(r;n_x,n_y,n_z,n_t)$ can be done 
in different ways. One possibility is using
the recursion relations of the previous section and the 
expression for the divergent part of $\BB(r)$ reported in
\reff{divergenzabosonico}. 
One can also attack the problem directly. Indeed if we define constants
$b_i(n_x,n_y,n_z,n_t)$ by the asymptotic expansion 
\be
\prod_{i=\{x,y,z,t\}}\left[{d^{n_i} \over d\alpha^{n_i}} 
(e^{-\alpha} I_0(\alpha)) \right]
=\, (-\alpha)^{-n_x-n_y-n_z-n_t} 
   \sum_{i=0}^\infty {b_i(n_x,n_y,n_z,n_t)\over \alpha^{i+2} }
\label{bxyzt}
\ee
and $q = r - n_x-n_y-n_z-n_t$ the divergent part of $\BB(r;n_x,n_y,n_z,n_t)$
(of course for $q\ge 2$) is given by 
\begin{eqnarray}
&&\sum_{i=2}^{q-1} b_{i-2}(n_x,n_y,n_z,n_t) 
    {\Gamma(q-i)\over 2^i \Gamma(p)} {1\over m^{2 q - 2 i} }\,
     \left(1 - \delta \log m^2 - \delta \sum_{k=q-i}^{p-1} {1\over k}\right)
     \nonumber \\
&& \qquad - b_{q-2} (n_x,n_y,n_z,n_t) {1\over 2^q \Gamma(p)} 
   \log m^2\left( 1 - {\delta\over2} \log m^2 - 
                       \delta \sum_{k=1}^{p-1} {1\over k} \right)
\end{eqnarray}
{}From this expression
we immediately see that the divergent part of $\FF(p,q)$ has the 
generic form
\be
{\cal D}^{(1)}(m) (1 -\delta \log m^2) + \delta {\cal D}^{(2)}(m) +\,
  \hbox{\rm log terms}\, + O(\delta^2)
\label{divergenzafermi2}
\ee
where ${\cal D}^{(1)}(m) $ and ${\cal D}^{(2)}(m)$ are polynomials 
in $1/m^2$ whose coefficients are rational numbers multiplied 
by $1/\pi^2$ and ``log terms" indicates terms which diverge as 
a power of $\log m^2$.

Exactly as in the bosonic case,
the knowledge of the divergent part of $\FF(p,q)$ can be used to simplify 
\reff{primariduzione}: indeed whenever an integral is multiplied 
by the infrared regulator we can substitute it with its divergent part.
Thus we can rewrite \reff{primariduzione} as 
\be
\FF (p,q;n_x,n_y,n_z,n_t) \,=\,
  \sum_{r=p-k+1}^p \sum_{s=q-k}^{q+k} a_{rs}(0,\delta) \FF(r,s) +\,
  {\cal R}(m,\delta)+\, O(m^2)
\label{primariduzionemassless}
\ee
If $p>0$ and $q\le 0$, as only terms with $s\le 0$ can appear in 
\reff{primariduzione}, the only $\FF(r,s)$ that can contribute to 
${\cal R}(m,\delta)$ have $r\ge 3$. Since $a_{rs}(m,\delta)$
is finite for $\delta\to 0$, we see that ${\cal R}(m,\delta)$
is a polynomial in $1/m^2$, finite for $\delta\to 0$. For $p>0$ and 
$q>0$ also $\FF(r,s)$ with $r\le 0$ can contribute to 
${\cal R}(m,\delta)$. As $a_{rs}(m,\delta)$ can behave as 
$1/\delta$ we have (see \reff{divergenzafermi2}) 
\be
{\cal R}(m,\delta) = {1\over\delta} 
   \left(1 - \delta\log m^2\right) {\cal R}^{(1)}(m)+\,
   {\cal R}^{(2)}(m) + O(\delta)
\ee
where ${\cal R}^{(1)}(m)$ and 
${\cal R}^{(2)}(m)$ are polynomials in $1/m^2$ whose 
coefficients are rational numbers multiplied by $1/\pi^2$.

For $p\le 0$ and $q\le 0$ it is easy to see that
${\cal R}(m,\delta) = 0$ as only finite integrals appear in the r.h.s.
of \reff{primariduzione}
while for $p\le 0$ and $q>0$ we have
\be
{\cal R}(m,\delta) = 
   \left(1 - \delta\log m^2\right) {\cal R}^{(1)}(m)+\,\delta\,
   {\cal R}^{(2)}(m) + O(\delta^2)
\ee
Thus in all cases the function ${\cal R}(m,\delta)$ has the form stated
in the introduction to the section.

\subsection{Second step: the identity ${\cal I}_1(p,q)$}

We will obtain here a new set of recursion relations using the identity 
${\cal I}_1(p,q)$.
Applying the 
previous recursion relations we can write each term in \reff{identity1}
as in \reff{primariduzionemassless} obtaining a non trivial relation of the 
form
\be
   \sum_{r,s} f_{rs}(p,q;\delta) \FF (r,s) +{\cal R}_\delta (p,q;m,\delta)
      = 0
\label{recursion1}
\ee
where $p-4\le r \le p$. 
Following our discussion of the bosonic case 
we will use \reff{recursion1} to obtain new recursion relations.
Let us first 
notice that in \reff{recursion1} there is only one term with 
$r=p-4$. It has $s=q+4$ and 
\be
f_{p-4,q+4}(p,q;\delta) = - {32 (q+1) (q+2) (q+3) \over 
        (p + \delta-1) (p + \delta - 2)(p + \delta - 3)}
\ee
Thus, if $q\not= -1,-2,-3$ we can solve
\reff{recursion1} in terms of $\FF(p-4,q+4)$. Shifting 
$p\to p+4$ and $q\to q-4$ we can express $\FF(p,q)$, $q\not= 1,2,3$ in terms
of $\FF(r,s)$ with $p+1\le r\le p+4$. We can then use this relation 
to express all integrals $\FF(p,q)$, $p\le -1$, $q\not= 1,2,3$ in
terms of $\FF(r,s)$ with either $0\le r \le 3$, $s$ arbitrary or
$1\le s \le 3$ and $p\le -1$.

Two observations are in order:
\begin{enumerate}
\item A careful analysis of the recursion shows that in the result 
      the coefficients of 
      $\FF (r,s)$ with $r=1,2,3$ are of order $\delta$.
      This property is very important: indeed it guarantees that
      when substituting these expressions in \reff{primariduzionemassless} 
      the coefficients
      of $\FF (p,q)$ with positive $p$ are finite for $\delta\to0$.
      This property would not be true if we were
      using the relation to eliminate also $\FF (r,s)$ with $r\ge 0$.
\item If we are considering $\FF(p,q)$ with $q\le 0$ then the result 
     is expressed only in terms of $\FF(r,s)$ with $0\le r\le 3$ and 
     $s\le 0$. 
\end{enumerate}
We could also try to 
use the same identity to obtain recursion relations which express
$\FF(p,q)$ in terms of $\FF(r,s)$ with $r<p$. To do this we should 
try to solve the identity for the $\FF(r,s)$ with the highest value of 
$r$, namely $r=p$. However in this case there are three terms with 
$r=p$, namely $\FF(p,q-2)$, $\FF(p,q-1)$ and $\FF(p,q)$
and thus we cannot obtain a recursion which decreases the value of $p$.
We will thus proceed in a different way.
We will solve the identity for 
$\FF(p,q-2)$, shifting $q\to q+2$. This is always possible 
as $f_{p,q-2}(p,q;\delta) = - 1$. In this way 
we obtain a recursion relation
which  expresses $\FF(p,q)$ in terms of $\FF(r,s)$ with $r<p$ and 
$\FF(p,s)$ with $s>q$. We will use this recursion to eliminate 
recursively all the integrals 
with $q\le -3$ and $p\ge 4$. The choice of stopping at $q=-3$ 
guarantees that only integrals $\FF(r,s)$ with $s\le 0$ are generated.

A third recursion relation can finally be obtained by solving
\reff{recursion1} in terms of $\FF(p,q)$.
This is also always possible as 
$f_{p,q} (p,q;\delta) = 256$.
In this way we can eliminate all integrals $\FF(p,q)$ with 
$p\ge4$ and $q>0$.

In conclusion, using the identity ${\cal I}_1 (p,q)$  we can rewrite every
$\FF(p,q)$ as 
\be
\FF(p,q) = \sum_{rs} c_{rs}(p,q;\delta) \FF(r,s) + 
   {\cal S}(p,q;m,\delta)
\ee
where in the sum we have either $0\le r\le 3$, 
$s$ arbitrary, or $r\le -1$ and $s=1,2,3$ or $r\ge 4$ and $s=-2,-1,0$.
It is easy to see that the properties mentioned at the beginning of 
the section are satisfied by $c_{rs}(p,q;\delta)$ and ${\cal S}(p,q;m,\delta)$.

\subsection{Third step: the identity ${\cal I}_2(p,q)$}

Let us now obtain a new set of recursion relations which allow to reduce
the remaining $\FF(p,q)$ in terms of a finite set of integrals.
We will use here the second identity ${\cal I}_2(p,q)$.
We will discuss separately four different regions:
\begin{enumerate}
\item $q\le 0$, $0\le p \le 3$;
\item $q> 0$, $0\le p \le 3$;
\item $q=1,2,3$, $ p \le -1$;
\item $q=-2,-1,0$, $ p \ge 4$.
\end{enumerate}

\subsubsection{The strip $q\le 0$, $0\le p \le 3$}

Let us first consider the integrals $\FF(p,q)$ with $q\le 0$ and $0\le p\le3$.

We start from ${\cal I}_2 (2,q+4)$. 
We can use the previous relations to obtain
an identity which involves only $\FF(r,s)$ with $0\le r \le 3$.
If $q\le -5$ we can solve it for $\FF(3,q)$ expressing it 
in terms of $\FF(r,s)$ with $r\le 2$ or $r=3$ and $s> q$. 
Notice that by stopping at $q=-5$ all integrals are expressed in terms 
of $\FF(r,s)$ with $s\le 0$.

We want now to obtain a relation for $\FF(2,q)$. In this case 
we start from ${\cal I}_2 (1,q+4)$.
Then we can use the relations of the first two steps
to obtain a recursion relation which involves only
$\FF(r,s)$, $0\le r \le 3$. Then we use the previous relation to eliminate
$\FF(3,q-2)$, $\FF(3,q-1)$ and 
$\FF(3,q)$. At the same time also $\FF(2,q-2)$
and $\FF(2,q-1)$ cancel. If $q\le -5$, this relation can then be solved 
in terms of $\FF(2,q)$, the result containing only 
$\FF(r,s)$ with $s\le 0$. 

In a completely analogous way we can derive recursions for 
$\FF(1,q)$ and $\FF(0,q)$. In the first case we start from 
${\cal I}_2 (0,q+6)$. We first apply the step-one and step-two 
relations, then use the previous relations to eliminate 
$\FF(3,q-2)$, $\FF(3,q-1)$, 
$\FF(3,q)$ and $\FF(2,q)$. If $q\le - 5$ we solve for $\FF(1,q)$.
Finally starting from
${\cal I}_2 (-1,q+6)$, eliminating 
$\FF(3,q-4)$, $\FF(3,q-3)$, $\FF(3,q-2)$,
$\FF(2,q-2)$ and $\FF(1,q-2)$ we get a relation 
for $\FF(0,q)$ valid for $q\le -5$. 

Using iteratively these four relations we are now able to express
every $\FF(p,q)$, $0\le p\le 3$, $q \le -5$, 
in terms of $\FF(r,s)$ with $0\le r \le 3$,
$-4 \le s \le 0$. 

\subsubsection{The strip $q>0$, $0\le p\le 3$}

Here we want to obtain relations analogous to the previous ones
but which {\em decrease} the value of $q$. The procedure is identical to the 
previous one.

We first consider ${\cal I}_2(2,q+1)$ and we use the step-one and step-two
relations to obtain an identity involving only 
$\FF(r,s)$, $0\le r\le 3$, then we solve for $\FF(3,q)$. We use this relation
for $q>0$. 

Analogously from ${\cal I}_2(1,q+1)$ and 
${\cal I}_2 (0,q+1)$ we get relations for $\FF(2,q)$ and 
$\FF(1,q)$ respectively valid for $q>2$ and $q> 4$. Finally
we consider ${\cal I}_2(-1,q+1)$: we apply the step-one and step-two
substitutions, then we use the previous relations 
to eliminate $\FF(3,q-3)$, $\FF(2,q-2)$ and $\FF(1,q-1)$ and finally
solve for $\FF(0,q)$. This relation is valid for $q>6$.
In this way we express all $\FF(p,q)$, $0\le p\le 3$, $q>0$ in
terms of $\FF(r,s)$, $0\le r\le 3$, $s\le 2(3-r)$.

\subsubsection{The strip $p\le -1$, $q=1,2,3$}

Here we consider ${\cal I}_2(p+3,1)$, apply the step-one and 
step-two relations to obtain an identity involving only
$\FF(r,s)$ with $s\le 3$, then we solve for $\FF(p,3)$. This
relation is valid for $p \le -1$. In the same way starting from 
${\cal I}_2(p+2,2)$ we get a relation for $\FF(p,2)$,
$p\le - 2$. Finally starting from ${\cal I}_2(p+1,3)$, after
eliminating $\FF(p-1,3)$, we get an identity for 
$\FF(p,1)$, $p\le -1$. Thus all integrals in this region but
$\FF(-1,2)$ can be rewritten in terms of $\FF(r,s)$, $r\ge 0$.

\subsubsection{The strip $p\ge 4$, $q=-2,-1,0$}

Here we start from ${\cal I}_2 (p-1,0)$. We apply the step-one
relations and then the step-two relations to eliminate
$\FF(p,-4)$ and $\FF(p,-3)$, then we solve for $\FF(p,-2)$.
This relation is valid for all $p\ge 4$.
Analogously, starting from ${\cal I}_2(p-1,-1)$, eliminating
$\FF(p,r)$, $-5\le r\le -2$, we get a relation for 
$\FF(p,-1)$, $p\ge 4$. Finally starting from ${\cal I}_2(p,-2)$,
eliminating $\FF(p+1,r)$, $-6\le r\le -1$ and $\FF(p,r)$,
$-4 \le r\le -1$ we get a relation for
$\FF(p,0)$, $p\ge 4$. In this way all $\FF(p,q)$ in this strip get
rewritten in terms of $\FF(r,s)$, $0\le r\le 3$, $s\le 0$.

\subsection{The fourth step: the last relations} \label{sec3.4}

In the preceding step we showed that all integrals can be rewritten
in terms of a finite number of them. However 
we have not used the identities ${\cal I}_1(p,q)$ and 
${\cal I}_2(p,q)$ for all possible values of $p$ and $q$.
For instance we never used ${\cal I}_1(p,q)$ for 
$p \le 3$ and $q=-1,-2,-3$ or $p \ge 4$ and $q=0$; much larger 
is the number of cases where the second identity has not been used.
We thus checked systematically if there were other values 
of $p,q$ for which the two identities were not trivially satisfied,
thus providing relations which could be used to further 
decrease the number of independent integrals. Using 
${\cal I}_1(p,q)$ with $(p,q)$ 
getting the values $(3,-1)$, $(2,-1)$,
$(1,-1)$, $(0,-1)$, $(3,-2)$, $(2,-2)$, $(1,-2)$, $(1,-3)$ and 
${\cal I}_2(2,0)$ , ${\cal I}_2 (3,-2)$ we obtain relations for 
$\FF(3,[0,-1])$, $\FF(2,[-3,-4])$, $\FF(1,[-3,-4])$ and 
$\FF(0,[-1,-2,-3,-4])$. 
These relations are reported in the appendix.
Notice that each integral is expressed in terms of $\FF(r,s)$ with $s\le 0$.
We stress that the identities we used to derive the relations 
were chosen arbitrarily; other choices could have been equally used.
However, once these relations have been computed we have found that 
${\cal I}_1(p,q)$, (resp. ${\cal I}_2(p,q)$),
is identically satisfied for all values of $q$ and for all $p<0$
(resp. $p\le 0$).

In a completely analogous way we have found that the identities
${\cal I}_1(4,0)$ and ${\cal I}_2(p,q)$ with 
$(p,q)$ getting the values $(3,2)$, $(3,3)$, $(1,2)$, $(1,3)$,
$(0,4)$, $(0,5)$ are not yet satisfied. We have used them to get relations
for $\FF(2,[1,2])$, $\FF(1,[3,4])$, $\FF(0,[4,5,6])$. 
They are reported in the appendix.

Collecting everything together we get
\be 
\FF(p,q;n_x,n_y,n_,n_t) = \sum_{rs} d_{rs}(\delta) \FF(r,s) + 
    {\cal T} (m,\delta)
\label{risultatofinale}
\ee
where $d_{rs}(\delta)$ and ${\cal T} (m,\delta)$ have the form explained at the 
beginning of the section.

Now, if $p>0$ and $q\le 0$ the second sum extends over ten values 
of $(r,s)$, i.e. $(0,0)$, $(1,t)$, $(2,t)$ and $(3,t+2)$ 
with $0\le t\le 2$. Let us consider the limit $\delta\to 0$. 
The polynomial ${\cal T} (p,q;m,\delta)$ is finite in this limit 
and the same is true for all $d_{rs}(p,q;\delta)$ with $r>0$. 
Thus the only coefficient which could have a $1/\delta$ divergence
is $d_{00}(p,q;\delta)$. However the result is finite for 
$\delta\to 0$ and thus also this coefficient is finite in this limit.
We can thus set simply $\delta=0$ and use the fact that 
$\FF(0,0) = 1 + O(\delta)$ to get 
\be
{\cal F}(p,q;n_x,n_y,n_,n_t) = \sum_{r>0;s} d_{rs}(p,q;0) {\cal F}(r,s) +
   d_{00}(p,q;0) + {\cal T} (p,q;m,0)
\ee
Finally let us consider the limit $m\to 0$. In this limit 
all ${\cal F}(r,s)$ appearing in the previous sum are finite 
except ${\cal F}(2,0)$ which we write as 
\be
   {\cal F}(2,0) =\, - {1\over 16 \pi^2}
   \left( \log m^2 + \gamma_E - F_0\right) + Y_0
\ee
where $Y_0$ is a numerical constant. Notice that if 
${\cal F}(p,q;n_x,n_y,n_,n_t) $ is finite, since ${\cal F}(2,0)$ 
is the only term which can contain $\log m^2$, 
$d_{2,0}(p,q;0) = 0$, i.e. the result depends only on the eight 
finite integrals. Numerical values are reported in Table 
\ref{costanti_fermioniche}.

\begin{table}
\begin{center}
\begin{tabular}{|c|c||c|c|}
\hline
${\cal F}(1,0)$ & \hphantom{$-$}  0.08539036359532067914  &
${\cal F}(1,-1)$ &                0.46936331002699614475  \\
${\cal F}(1,-2)$ & \hphantom{$-$} 3.39456907367713000586  &
${\cal F}(2,-1)$ &                0.05188019503901136636  \\
${\cal F}(2,-2)$ & \hphantom{$-$} 0.23874773756341478520  &
${\cal F}(3,-2)$ &                0.03447644143803223145  \\
${\cal F}(3,-3)$ & \hphantom{$-$} 0.13202727122781293085  &
${\cal F}(3,-4)$ &                0.75167199030295682254  \\
$Y_0$           & $-$             0.01849765846791657356  &
$Y_1$           &                 0.00376636333661866811  \\
$Y_2$           & \hphantom{$-$}  0.00265395729487879354  &
$Y_3$           &                 0.00022751540615147107  \\
\hline
\end{tabular}
\end{center}
\caption{Numerical values of the constants appearing in the fermionic 
integrals.
}
\label{costanti_fermioniche}
\end{table}

Let us now consider the case $p>0$ and $q>0$. In this case in the sum
on the r.h.s of \reff{risultatofinale} we can also have 
$\FF(-1,2)$, $\FF(0,1)$, $\FF(0,2)$, $\FF(0,3)$, $\FF(1,1)$ and 
$\FF(1,2)$; ${\cal T}(p,q;m,\delta)$ has the form
\be
{1\over \delta} {\cal T}^{(1)}(p,q;m) (1 - \delta \log m^2) + 
{\cal T}^{(2)}(p,q;m) + O(\delta)
\ee
where ${\cal T}^{(1)}(p,q;m) $ and ${\cal T}^{(2)}(p,q;m)$ are polynomials
in $1/m^2$. Let us consider the limit $\delta\to 0$. The only coefficients
that may behave as $1/\delta$ are those with $r\le 0$. Writing in this
case 
\be
d_{rs}(\delta) = {1\over \delta} d_{rs}^{(1)} + 
      d_{rs}^{(2)} + O(\delta)
\ee
the cancellation of the $1/\delta$ terms gives the equation
\be
d_{03}^{(1)} {\cal B}(3) + d_{02}^{(1)} {\cal B}(2) +
(d_{01}^{(1)} + 2 d_{-1,2}^{(1)}) {\cal B}(1) + d_{00}^{(1)}  +
{\cal T}^{(1)}(m) = 0
\label{unosudeltaequation}
\ee
where we have used $\FF(0,q) = {\cal B}(q) + O(\delta)$ and 
\be
\FF(-1,2) = {\cal B}(1) - {\cal B}(2;2) +1/4 + O(\delta) = 
2 {\cal B}(1) + O(\delta)
\ee
In all cases we have found that \reff{unosudeltaequation}
is satisfied in a trivial way, i.e.
\begin{eqnarray}
&& d_{03}^{(1)} = d_{02}^{(1)} = d_{00}^{(1)} = 0 \nonumber \\
&& d_{01}^{(1)} + 2 d_{-1,2}^{(1)} = 0   \nonumber \\
&& {\cal T}^{(1)}(m) = 0 
\label{relazioniperid}
\end{eqnarray}
This is not completely surprising: indeed cancellation of the terms
divergent for $m\to0$ requires $d_{03}^{(1)} = - 8 d_{02}^{(1)}$ and 
\be
      {\cal T}^{(1)}(m) = - {d_{03}^{(1)}\over 32 \pi^2 m^2} + {t\over\pi^2}
\ee
where $t$ is a rational number. Then \reff{unosudeltaequation} becomes
\be
{d_{03}^{(1)}\over 256} Z_1 + 
  ( d_{01}^{(1)} + 2 d_{-1,2}^{(1)} ) Z_0 + 
   d_{00}^{(1)} - {d_{03}^{(1)}\over 1024} + 
  {1\over \pi^2} \left( t - {13\over 1536} d_{03}^{(1)} \right) = 0
\ee
Thus, if for some $p,q,n_x,n_y,n_z,n_t$,
the relations \reff{relazioniperid}  were not satisfied, one would get an
equation which would allow to eliminate 
either $Z_0$ or $Z_1$.

Because of \reff{relazioniperid}, in the limit $\delta\to0$,
we can then substitute in 
\reff{risultatofinale} $\FF(0,3) = {\cal B}(3)$,
$\FF(0,2) = {\cal B}(2)$, $\FF(0,0) = 1$ and rewrite 
\be
 d_{01} \FF(0,1) + d_{-1,2} \FF(-1,2) \, =\, 
{1\over2} d_{01} (2 \FF(0,1) - \FF(-1,2)) + 
   (2 d_{-1,2} + d_{01}) {\cal B}(1) + O(\delta)
\ee
showing that $\FF(0,1)$ and $\FF(-1,2)$ appear in the result 
only in the fixed combination $2 \FF(0,1) - \FF(-1,2)$. 
In conclusion, for $q>0$ the result beside the nine integrals 
which appear for $q\le 0$ contains also the bosonic constants
$Z_0$, $Z_1$, $F_0 -\gamma_E$ and the integrals 
${\cal F}(1,1)$, ${\cal F}(1,2)$ and 
$\lim_{\delta\to0}(2 \FF(0,1) - \FF(-1,2))/\delta$. As in the bosonic case,
instead of these three quantities we have parameterized our results
in terms of three infrared-finite integrals. We introduce:
\begin{eqnarray}
Y_1 &=& {1\over 8}\, {\cal F} (1,1;1,1,1) \nonumber \\
Y_2 &=& {1\over 16}\, {\cal F} (1,1;1,1,1,1) \nonumber \\
Y_3 &=& {1\over 16}\, {\cal F} (1,2;1,1,1) 
\end{eqnarray}
Their numerical values are reported in Table \ref{costanti_fermioniche}.
The relation with the original integrals is:
\begin{eqnarray}
&& \lim_{\delta\to0} {1\over\delta} (2\, \FF(0,1) - \FF(-1,2))\, =\, 
  {1\over4} - 12\, Y_1 - 3\, Z_0 + 2\, {\cal F}(1,0) \\
&& {\cal F}(1,1) \, =\, - {1\over 16 \pi^2} (\log m^2 + \gamma_E - F_0)
    - {1\over 192} + {1\over 16\pi^2} + Y_0 - {1\over4} Y_1 + 
   {1\over 16} Y_2 \nonumber \\
&& \qquad + {1\over 768} {\cal F}(1,-2) + {1\over 192} {\cal F}(1,-1) + 
    {59\over 192} {\cal F}(1,0) + {1\over 768} {\cal F}(2,-2) -
    {25 \over 48} {\cal F}(2,-1) 
   \nonumber \\ [-3mm] 
&& {}     \\
&& {\cal F}(1,2) \, =\, {1\over 32 \pi^2 m^2} - {19\over 12288} + 
   {307\over 18432 \pi^2} +\, {3\over 64} Y_0 - {19\over 256} Y_1 
   + {19\over 1024} Y_2 + {1\over 8} Y_3 \nonumber \\
&& \qquad + {1\over 768} Z_0 + {19\over 49152} {\cal F}(1,-2) 
   + {53\over 36864} {\cal F}(1,-1) + {187\over 4096} {\cal F}(1,0) 
    \nonumber \\
&& \qquad + {5497\over 147456} {\cal F}(2,-2) - 
    {293\over 9216} {\cal F}(2,-1) - {35\over 6144} {\cal F}(3,-4)
   - {19\over 512} {\cal F}(3,-3) \nonumber \\ 
&& \qquad - {173\over 2304} {\cal F}(3,-2)
\end{eqnarray}
As in the bosonic case, let us notice that $\log m^2$ appears always in the 
fixed combination $(\log m^2 + \gamma_E - F_0)$. Thus in 
finite integrals $F_0 - \gamma_E$ does not appear. 

To conclude this section we want to add a few remarks on the numerical 
evaluation of the constants appearing in Table 
\ref{costanti_fermioniche}. A direct evaluation of the integrals
does not provide accurate results: we have thus used a different 
procedure inspired by the work of \cite{LW}. To evaluate the constants 
${\cal F}(p,q)$ with $q \le 0$ we have considered the integrals 
$J_q = {\cal F}(1,-q)$ with $6\le q\le 13$. To compute them 
we have first calculated the sums 
\be
    J_{q,L} = {1\over L^4} \sum_k {D_B(k,0)^q\over D_F(k,0)}
\label{estrapolazionenumerica}
\ee
where $k$ runs over the points $k = (2\pi/L) (n_1+\smfrac{1}{2},
n_2+\smfrac{1}{2},n_3+\smfrac{1}{2},n_4+\smfrac{1}{2})$,
$0\le n_i < L$ for various values of $L$ between 50 and 100. Then we tried
to extrapolate $J_{q,L}$ using the form 
\be 
       J_{q,L} =\, J_q \left(1 + {a\over L^{2q+2}} \right)
\ee
In all cases the correction turned out to be completely negligible.
In practice 
$J_q$ could be determined with a relative error $\ltapprox 10^{-25}$.
Then each $J_q$ was expressed in terms of the basic constants. We obtained in 
this way 8 equations which were solved giving the results of Table
\ref{costanti_fermioniche}. 

Analogously to compute $Y_0$, $Y_1$, $Y_2$ and $Y_3$, we computed numerically
${\cal F}(1,1,8)$, ${\cal F}(2,1,9)$, ${\cal F}(3,1,10)$ and 
${\cal F}(5,2,11)$ and then solved the corresponding equations.

\section{Integrals in other infrared regularizations}

In the preceding two sections we have discussed the computation of bosonic
and mixed bosonic-fermionic integrals using as infrared regulator 
a mass $m$. Here we want to discuss other types of infrared regularization:
first we will consider the dimensional regularization \cite{Kawai}
and then we will consider mixed bosonic-fermionic integrals 
with the exact Wilson-fermion propagator 
\reff{trueFermiprop}.

\subsection{Dimensional regularization}

In this case we consider the integrals \cite{Kawai}
\begin{eqnarray}
{\cal B}^{DR}(p;n_x,n_y,n_z,n_t) &= &
    \int^\pi_{-\pi} {d^d k\over (2 \pi)^d} 
    {\hat{k}_x^{2 n_x} \hat{k}_y^{2 n_y} \hat{k}_z^{2 n_z} \hat{k}_t^{2n_t} 
          \over D_B(k,0)^p} \label{BDR}\\
{\cal F}^{DR}(p,q;n_x,n_y,n_z,n_t) &= &
    \int^\pi_{-\pi} {d^d k\over (2 \pi)^d} 
    {\hat{k}_x^{2n_x} \hat{k}_y^{2n_y} \hat{k}_z^{2n_z} \hat{k}_t^{2n_t} 
          \over D_F(k,0)^p D_B(k,0)^q}  \label{FDR}
\end{eqnarray}
It is easy to see that the basic recursion relations 
\reff{bosonicrec1}, \reff{rec2}, \reff{fermionicrec1},
\reff{fermionicrec2}, \reff{fermionicrec3} can be easily generalized
to dimensionally-regularized integrals. The relation 
\reff{identitabosonica} and \reff{identity1}, \reff{identity2}
are instead intrinsically four-dimensional identities and for 
this reason we have made the computation using a mass as a 
regulator. Now, we will show how to compute \reff{BDR} and 
\reff{FDR} from their mass-regularized counterparts.

Let us begin with the bosonic case, considering 
${\cal B}(p;n_x,n_y,n_z,n_t)$. If $q\equiv p - n_x - n_y - n_x - n_t < 2$
the integral is finite and thus independent of the infrared regulator. For 
$q \ge 2$ let us rewrite
\be
\hskip -20pt 
{\cal B}(p;n_x,n_y,n_z,n_t) = \, 
   {1\over 2^p \Gamma(p)} \int_0^\infty d\alpha\, \alpha^{p-1} 
    e^{-\alpha m^2/2} \prod_{i=\{x,y,z,t\} } 
     \left[ (-2)^{n_i} {d^{n_i}\over d\alpha^{n_i} }
       ( e^{-\alpha} I_0 (\alpha) ) \right]
\ee
Then define
\begin{eqnarray}
\hskip -20pt
F(p;n_x,n_y,n_z,n_t) &=& {1\over 2^p \Gamma(p)} 
      \int^1_0 d\alpha\, \alpha^{p-1} \prod_{i=\{x,y,z,t\} }
     \left[ (-2)^{n_i} {d^{n_i}\over d\alpha^{n_i} }
       ( e^{-\alpha} I_0 (\alpha) ) \right] \nonumber \\
&& + {1\over 2^p \Gamma(p)}
      \int_1^\infty d\alpha\, \alpha^{p-1} \left\{ \prod_{i=\{x,y,z,t\} }
     \left[ (-2)^{n_i} {d^{n_i}\over d\alpha^{n_i} }
       ( e^{-\alpha} I_0 (\alpha) ) \right] \right. \nonumber \\
&& \qquad \qquad \left.
     - \left({2\over \alpha}\right)^{n_x+n_y+n_z+n_t} \,
       \sum_{i=0}^{q-2} {b_i(n_x,n_y,n_z,n_t)\over \alpha^{i+2} } 
      \right\}
\end{eqnarray}
where $b_i(n_x,n_y,n_z,n_t)$ are defined by \reff{bxyzt}.
Then for $m^2\to 0$ we have 
\begin{eqnarray}
&& {\cal B}(p;n_x,n_y,n_z,n_t) = F(p;n_x,n_y,n_z,n_t) \nonumber \\
&& \qquad + {1\over \Gamma(p)} 
   \sum_{i=2}^{q-1} b_{i-2}(n_x,n_y,n_z,n_t) 
   \left[ {\Gamma(q-i)\over 2^i} {1\over m^{2 q - 2 i} } - 
      {1\over 2^q} {1\over q-i} \right] \nonumber \\
&& \qquad - {1\over 2^q \Gamma(p)} b_{q-2} (n_x,n_y,n_z,n_t) 
   \left( \log {m^2\over 2} + \gamma_E \right)
\end{eqnarray}
Analogously we have 
\be
\hskip -20pt 
{\cal B}^{DR}(p;n_x,n_y,n_z,n_t) = \, 
   {1\over 2^p \Gamma(p)} \int_0^\infty d\alpha\, \alpha^{p-1} 
   \left( e^{-\alpha} I_0 (\alpha) \right)^{d-4} \prod_{i=\{x,y,z,t\} } 
     \left[ (-2)^{n_i} {d^{n_i}\over d\alpha^{n_i} }
       ( e^{-\alpha} I_0 (\alpha) ) \right] 
\ee
so that for $\epsilon\equiv 4 - d\to 0$ we have
\be
- {2\over \epsilon} {1\over 2^q \Gamma(p)} 
   b_{q-2} (n_x,n_y,n_z,n_t) \left(1 + \, \epsilon 
   c_{q-2} (n_x,n_y,n_z,n_t)\right) + F(p;n_x,n_y,n_z,n_t)
\ee
where $c_i(n_x,n_y,n_z,n_t)$ are defined by
\begin{eqnarray}
&& \prod_{i=\{x,y,z,t\}} \left[
     (-\alpha)^{n_i} 
     {d^{n_i}\over d\alpha^{n_i} } (e^{-\alpha} I_0 (\alpha) )\right] 
     \log(e^{-\alpha} I_0(\alpha) \sqrt{\alpha}) 
\nonumber \\
&& \qquad \quad =\, - \sum_{i=0}^\infty {1\over \alpha^{i+2}} 
    b_i (n_x,n_y,n_z,n_t) c_i(n_x,n_y,n_z,n_t)
\end{eqnarray}
Then, by comparison, we have
\begin{eqnarray}
&& \hskip -40pt 
   {\cal B}^{DR} (p;n_x,n_y,n_z,n_t)\, =\, {\cal B}(p;n_x,n_y,n_z,n_t) 
\nonumber \\
&& - {1\over \Gamma(p)} 
    \sum_{i=2}^{q-1} b_{i-2} (n_x,n_y,n_z,n_t) 
    \left( {\Gamma(q-i)\over 2^i} {1\over m^{2q - 2 i}} - 
        {1\over 2^q(q-i)} \right) \nonumber \\
&& - {1\over 2^q \Gamma(p)} b_{q-2} (n_x,n_y,n_z,n_t) 
   \left(-\log {m^2\over2} - \gamma_E + {2\over \epsilon} 
       + 2 c_{q-2} (n_x,n_y,n_z,n_t) \right)
\end{eqnarray}
A simplification occurs if the integral is logarithmically infrared 
divergent, i.e. if $q=2$. In this case, as $c_0(n_x,n_y,n_z,n_t) = 
\smfrac{1}{2} \log 2\pi$, we have
\begin{eqnarray}
&& \hskip -40pt 
   {\cal B}^{DR} (p;n_x,n_y,n_z,n_t)\, =\, {\cal B}(p;n_x,n_y,n_z,n_t) 
\nonumber \\
&& - {1\over 4 \Gamma(p)} b_0 (n_x,n_y,n_z,n_t) 
    \left [ - \log m^2 - \gamma_E + {2\over \epsilon} + \log 4 \pi\right]
\end{eqnarray}
Thus, for these integrals, we can use a very simple recipe to go from
the mass regularization to the dimensional one: simply substitute in
each integral $\log m^2 + \gamma_E$ with $2/\epsilon + \log 4 \pi$.

Let us now consider the fermionic case. Again we should consider 
only the case $Q \equiv p + q - n_x - n_y - n_z - n_t \ge 2$. 
Then let us rewrite 
\begin{eqnarray}
&&{\cal F}^{DR}(p,q;n_x,n_y,n_z,n_t) = \int {d^dk\over (2\pi)^d} 
    \left[ {
    \hat{k}^{2n_x}_x \hat{k}^{2n_y}_y \hat{k}^{2n_z}_z \hat{k}^{2n_t}_t
    \over D_F(k,0)^{p} D_B(k,0)^q} \right.  \nonumber \\
    && \qquad \qquad  \left. - \, 
    {\hat{k}^{2n_x}_x \hat{k}^{2n_y}_y \hat{k}^{2n_z}_z \hat{k}^{2n_t}_t
         \over D_B(k,0)^{p+q}} 
          \sum_{l=0}^{Q-2} {-p\choose{l}}
           \left( {\Delta D_F(k,0)\over D_B(k,0)}\right)^l \right]
       \nonumber \\
     && \qquad + \sum_{l=0}^{Q-2} {-p\choose{l}} 
        \, \int {d^d k\over (2\pi)^d} 
       \hat{k}^{2n_x}_x \hat{k}^{2n_y}_y \hat{k}^{2n_z}_z \hat{k}^{2n_t}_t
       { \Delta D_F(k,0)^l\over D_B(k,0)^{p+q+l} }
\end{eqnarray}
where $\Delta D_F(k,m) \equiv D_F(k,m) - D_B(k,m)$. The first integral in the 
is clearly finite. We can thus set $\epsilon = 0$. Then it is easy to see
that we can add everywhere a mass, i.e. substitute 
$D_B(k,0)$ with $D_B(k,m)$ and analogously for $D_F(k,0)$, without changing
its value in the limit $m\to 0$. Thus we get finally
\begin{eqnarray}
&&{\cal F}^{DR}(p,q;n_x,n_y,n_z,n_t) =
  {\cal F} (p,q;n_x,n_y,n_z,n_t) \nonumber \\
&& + \sum_{l=0}^{Q-2} {-p\choose{l}} \left\{
        \, \int {d^d k\over (2\pi)^d}
      \hat{k}^{2n_x}_x \hat{k}^{2n_y}_y \hat{k}^{2n_z}_z \hat{k}^{2n_t}_t
       { \Delta D_F(k,0)^l\over D_B(k,0)^{p+q+l} } \right.
\nonumber \\
&& \qquad \qquad \left. 
     -\, 
    \int {d^4k\over (2\pi)^4}
       \hat{k}^{2n_x}_x \hat{k}^{2n_y}_y \hat{k}^{2n_z}_z \hat{k}^{2n_t}_t
       { \Delta D_F(k,m)^l\over D_B(k,m)^{p+q+l} } \right\}
\end{eqnarray}
In the last term purely bosonic integrals are involved and we have already 
discussed how to compute the difference between their value in the 
two regularizations. Notice that, as in the bosonic case, logarithmically 
divergent integrals can be dealt with easily: simply substitute 
$(2/\epsilon + \log 4 \pi)$ to $(\log m^2 + \gamma_E)$.

\subsection{Massive Wilson-fermion-propagator integrals}

Here we want to consider integrals of the form
\be
\widehat{{\cal F}}(p,q;n_x,n_y,n_z,n_t) = 
    \int^\pi_{-\pi} {d^4 k\over (2 \pi)^4} 
    {\hat{k}_x^{2n_x} \hat{k}_y^{2n_y} \hat{k}_z^{2n_z} \hat{k}_t^{2n_t} 
          \over \widehat{D}_F(k,m)^p D_B(k,m)^q}  
\ee
where $\widehat{D}_F(k,m)$ is defined in \reff{trueFermiprop}. 
When $Q\equiv p + q - n_x - n_y - n_z - n_t \ge 2$ the integrals 
diverge for $m\to 0$. We will now relate them to ${\cal F}$. Indeed 
we can rewrite
\begin{eqnarray}
&& \widehat{{\cal F}}(p,q;n_x,n_y,n_z,n_t) = \,
    {\cal F} (p,q;n_x,n_y,n_z,n_t) \nonumber \\
&& \qquad + \sum_{l=1}^{2 Q - 4} {-p \choose l}
  \int^\pi_{-\pi} {d^4 k\over (2 \pi)^4} 
 {\hat{k}_x^{2n_x} \hat{k}_y^{2n_y} \hat{k}_z^{2n_z} \hat{k}_t^{2n_t}
   \Delta \widehat{D}_F(k,m)^l \over 
  D_F(k,m)^{p+l} D_B(k,m)^q} +\, O(m)
\label{Fhat}
\end{eqnarray}
where $\Delta \widehat{D}_F(k,m) \equiv
     \widehat{D}_F(k,m) - D_F(k,m) = m \hat{k}^2$.

Let us notice that if the integral is logarithmically divergent 
$(Q=2)$, for $m\to 0$ we have $\widehat{{\cal F}} = {\cal F}$.
For integrals with $Q>2$ we see from the explicit expression 
\reff{Fhat} the the divergent part is now a polynomial 
in $1/m$ instead of $1/m^2$. For this reason 
the expression for $\widehat{{\cal F}}$ are in general 
more cumbersome than those involving ${\cal F}$ and this is why
we have studied integrals with \reff{Fermiprop} 
instead of \reff{trueFermiprop}.

\section {Applications}

In this section, as an application of our method we will give a few
examples. In the first subsection we will report analytic expressions
for various renormalization constants whose value is reported in the
literature only in numerical form or is expressed in terms of 
cumbersome integrals. Then we will briefly discuss the computation of 
the fermionic propagator.

\subsection{Analytic expressions}

\subsubsection{Fermionic self-energy}
We want to give here the expression for the fermionic self-energy.
The first computation for the Wilson action in Feynman gauge 
was given in \cite{Gonzalez-Yndurain-Martinelli}
and it was subsequently corrected in \cite{Hamber-Wu}\footnote{
Notice however that formula (3.15) in \cite{Hamber-Wu} contains a 
misprint: the correct result is given in formula (10b) of 
\cite{Gonzalez-Yndurain-Martinelli}.}. The fermionic self-energy at one loop
has the generic form 
\be
\Sigma^{LATT}(p^2,m^2)\, =\, g^2 {N^2-1\over 2N} 
\left (\delta m + i {\not \mathrel{\,{\rm p}}} \Sigma_1 (p^2,m^2) + 
    m \Sigma_2 (p^2,m^2)\right)
\ee
For $r_W=1$, in Feynman gauge, in terms of our basic integrals we have
\begin{eqnarray}
\delta m &=& - Z_0 - 2 {\cal F}(1,0) \approx - 0.3257141174
\\
\Sigma_1 (p^2,m^2) &=& {1\over 16\pi^2} (2 G(p^2a^2,m^2 a^2) + \gamma_E - F_0)
   + {1\over 8} Z_0 + {1\over 192} - {1\over 32 \pi^2} - Y_0 + 
     {1\over 4} Y_1 
\nonumber \\
&& - {1\over 16}Y_2 + 12\, Y_3 - {1\over 768} {\cal F}(1,-2) - 
   {1\over 192} {\cal F}(1,-1) + {109\over 192} {\cal F}(1,0) 
\nonumber \\ 
&& - {1\over 768} {\cal F}(2,-2) + {25\over 48} {\cal F}(2,-1) 
\nonumber \\ 
&\approx&
  {1\over 8\pi^2} G(p^2a^2,m^2 a^2) + 0.0877213749 
\\
\Sigma_2 (p^2,m^2) &=&
{1\over 4\pi^2} (F(p^2a^2,m^2 a^2) + \gamma_E - F_0) + 
    {1\over 48} - {1\over 4\pi^2} - 4\, Y_0 +\, Y_1 - {1\over 4} Y_2 
\nonumber \\
&& -{1\over 192} {\cal F}(1,-2) - {1\over 48} {\cal F}(1,-1) -
   {83\over 48} {\cal F}(1,0) - {1\over 192} {\cal F}(2,-2) 
\nonumber \\ 
&& + 
   {49\over 12} {\cal F}(2,-1) \approx 
   {1\over 4\pi^2} F(p^2 a^2,m^2 a^2) + 0.0120318529
\end{eqnarray}
where 
\begin{eqnarray}
F(p^2a^2,m^2 a^2) &=& \int_0^1 dx \log[(1-x) (p^2 x + m^2) a^2]  \\
G(p^2a^2,m^2 a^2) &=& \int_0^1 dx \, x \log[(1-x) (p^2 x + m^2) a^2] 
\end{eqnarray}

\subsubsection{Gluonic self-energy}

Let us now consider the gluonic self-energy which was firstly computed
in \cite{Weisz-self,Kawai}. 
The contribution of the fermions, for $r_W=1$, can be 
expressed in terms of our basic integrals as
\be 
\Pi^f_{\mu\nu}(p,m) \, =\, 
 {N_f\over2} g^2 (p^2 \delta_{\mu\nu} - p_\mu p_\nu )
\left[ {1\over 12\pi^2} \left (6\ H(p^2a^2,m^2 a^2) + \gamma_E 
    - \log 4 \pi 
     \right) + L \right]
\ee
where $N_f$ is the number of fermions which are in the fundamental 
representation of $SU(N)$, $m$ the fermion mass (for simplicity
we assume all fermions to have the same mass), 
\begin{eqnarray}
L &=& - {1\over9} - {1\over 12\pi^2} (F_0 - \log 4 \pi) - 
   {4\over 3} Y_0 + {1\over36} {\cal F} (1,-2) \nonumber \\
  && \qquad + {1\over 18} {\cal F}(1,-1) - {7\over6} {\cal F}(1,0) 
   + {5\over 24} {\cal F}(2,-2) + {2\over3} {\cal F}(2,-1)
\end{eqnarray}
and 
\be
H(p^2 a^2, m^2 a^2) \, =\, 
\int_0^1 dx\; x(1-x)\, \log[ x(1-x) p^2 a^2 + m^2 a^2]
\ee
Numerically $L\approx 0.0031048512$ in agreement with \cite{Kawai}.

For $p\to 0$ we have
\be
\Pi^f(p,m) =\, {N_f\over2} g^2 (p^2 \delta_{\mu\nu} - p_\mu p_\nu)
\left[ {1\over 12 \pi^2} \log m^2 - 0.013391999 + O(p^2)\right]
\ee
while for $m=0$ we have
\be
\Pi^f(p,0) = \, {N_f\over2} g^2 (p^2 \delta_{\mu\nu} - p_\mu p_\nu)
\left[ {1\over 12 \pi^2} \log p^2 - 0.027464385 \right]
\ee

\subsubsection{Renormalization constants for bilinear fermion 
operators with the clover action}

We want to give here the renormalization constants 
of bilinear fermion operators with the {\em clover} action
\cite{clover-action,clover-Roma}, using the explicit expressions in
terms of lattice integrals of
\cite{bilinearfermions}. We define local operators 
\be
{\cal O}^{LATT,loc}_\Gamma(x) \,=\,
 \overline{\psi}(x)\Gamma \psi(x) 
\label{eq5.10}
\ee
and improved operators 
\be
{\cal O}^{LATT,imp}_\Gamma(x) \,=\,
 \overline{\psi}(x)\Gamma \psi(x) + \,
  {r_W\over2} \sum_\mu 
      \left(D_\mu \overline{\psi}(x) \gamma_\mu \Gamma \psi(x) 
     - \overline{\psi}(x) \Gamma \gamma_\mu D_\mu \psi(x) \right)
\label{eq5.11}
\ee
where $\Gamma$ is a Dirac matrix and 
\be
D_\mu \psi(x) = {1\over2} \left[ 
    U_\mu(x) \psi (x +\mu) - U^{+}_\mu (x-\mu) \psi(x-\mu)\right]
\ee
For each operator we compute a finite renormalization constant
$Z$ such that 
\be
\< f | {\cal O}^{CONT}(x) | i \> \, =\, 
Z \< f | {\cal O}^{LATT}(x) | i \>
\ee
where $f$ and $i$ are arbitrary external states. In the continuum 
we adopt the $\overline{MS}$-scheme with scale $\mu =1/a$.
We write at one loop
\be
Z = 1 + g^2 {N^2-1\over 8 N} \Delta Z
\ee
Expressions for $\Delta Z$, in Feynman gauge, are reported in 
\cite{bilinearfermions} in terms of quite complicated integrals.
The expression for $r_W=1$ in terms of our basic integrals is reported in 
Table \ref{Z_bilineari_local} and 
Table \ref{Z_bilineari_improved}. The final numerical values are in agreement
with those of \cite{bilinearfermions}.

\renewcommand{\arraystretch}{1.5}
\begin{table}
\begin{center}
\begin{tabular}{|l|rrrrr|}
\hline
 & \multicolumn{1}{r}{$\Delta Z_{Id}$}   &  
   \multicolumn{1}{r}{$\Delta Z_{\gamma_5}$} & 
   \multicolumn{1}{r}{$\Delta Z_{\gamma_\mu}$} &
   \multicolumn{1}{r}{$\Delta Z_{\gamma_5\gamma_\mu}$}   &   
   \multicolumn{1}{r|}{$\Delta Z_{\sigma_{\mu\nu}}$ }     \\
\hline
$1$ & $\smfrac{31}{48}$ & $\smfrac{7}{48}$ 
         & $\smfrac{3}{32}$    &    $\smfrac{11}{32}$            
         & $\smfrac{23}{144}$         \\
$1/\pi^2$ & $- \smfrac{223}{128}$ & $-\smfrac{139}{128}$  
         & $-\smfrac{19}{64}$      & $-\smfrac{5}{8}$   
         & $\smfrac{73}{384}$     \\
$ (F_0 - \gamma_E)/4 \pi^2 $ & $-3$ & $-3$ & 
         $0$       & $0$       & $1$ \\          
$Y_0$ 
         & $- 12$ & $- 12$ & $- 2$ & $- 2$                
         & $\smfrac{4}{3}$  \\
$Y_1$
         & $- 18$ & $-6$    
         & $-\smfrac{23}{2}$       &  $-\smfrac{35}{2}$      
         & $-\smfrac{46}{3}$     \\
$Y_2$
         & $- \smfrac{1}{4}$ & $- \smfrac{1}{4}$ 
         &  $ \smfrac{3}{8}$ & $ \smfrac{3}{8}$ 
         & $ \smfrac{7}{12}$ \\
$Y_3$
        & $- 48$              &$- 48$              
        & $- 48$              &$- 48$              
        & $- 48$              \\
$Z_0$
        & $6$              & $2$              
        & $3$              & $5$              
        & $4$             \\      
$Z_1$
        & $- \smfrac{3}{4}$ & $- \smfrac{3}{4}$            
        & $- \smfrac{3}{4}$ & $- \smfrac{3}{4}$            
        & $- \smfrac{3}{4}$             \\      
${\cal F}(1,-2)$
        & $-\smfrac{5}{96}$ & $\smfrac{7}{96}$
        & $\smfrac{15}{128}$ & $\smfrac{7}{128}$ 
        & $\smfrac{1}{9}$ \\
${\cal F}(1,-1)$
        & $-\smfrac{391}{192}$ & $-\smfrac{139}{192}$
        & $-\smfrac{39}{32}$ & $-\smfrac{15}{8}$
        & $-\smfrac{923}{576}$ \\
${\cal F}(1, 0)$
        & $-\smfrac{53}{8}$ & $-\smfrac{99}{8}$ 
        & $-\smfrac{1031}{96}$ & $-\smfrac{755}{96}$
        & $-\smfrac{665}{72}$ \\
${\cal F}(2,-2)$
        & $\smfrac{65}{192}$ & $-\smfrac{271}{192}$
        & $-\smfrac{235}{128}$ & $-\smfrac{123}{128}$
        & $-\smfrac{971}{576}$ \\
${\cal F}(2,-1)$
        & $-\smfrac{407}{8}$ & $\smfrac{117}{8}$
        & $\smfrac{631}{24}$ & $-\smfrac{155}{24}$
        & $\smfrac{1387}{72}$   \\
${\cal F}(3,-3)$
        & $\smfrac{77}{32}$ & $\smfrac{105}{32}$
        & $\smfrac{7}{2}$ & $\smfrac{49}{16}$
        & $\smfrac{329}{96}$ \\
${\cal F}(3,-2)$
        & $\smfrac{1057}{16}$ & $-\smfrac{203}{16}$
        & $-\smfrac{259}{8}$ & $7$
        & $-\smfrac{413}{16}$  \\
Total
        & $-0.4891266$ & $-0.5669576$
        & $-0.3882898$ & $-0.3493743$
        & $-0.2819883$      \\
\hline
Ref. \protect\cite{bilinearfermions} 
        & $-0.495$ & $-0.573$ & $-0.388$ & $-0.349$ & $-0.280$ \\
\hline
\end{tabular}
\end{center}
\caption{Results for $\Delta Z_\Gamma$ for the local operators
\protect\reff{eq5.10} : we report here the coefficients 
of the various constant appearing in the result. ``Total" is 
the numerical value of $\Delta Z_\Gamma$.}
\label{Z_bilineari_local}
\end{table}

\begin{table}
\begin{center}
\begin{tabular}{|l|rrrrr|}
\hline
 & \multicolumn{1}{r}{$\Delta Z_{Id}$}   &  
   \multicolumn{1}{r}{$\Delta Z_{\gamma_5}$} & 
   \multicolumn{1}{r}{$\Delta Z_{\gamma_\mu}$} &
   \multicolumn{1}{r}{$\Delta Z_{\gamma_5\gamma_\mu}$}   &   
   \multicolumn{1}{r|}{$\Delta Z_{\sigma_{\mu\nu}}$ }     \\
\hline
$1$ & $-\smfrac{151}{48}$ & $\smfrac{7}{48}$ 
         & $\smfrac{19}{24}$    &    $-\smfrac{41}{48}$            
         & $\smfrac{11}{24}$         \\
$1/\pi^2$ & $- \smfrac{279}{128}$ & $-\smfrac{139}{128}$  
         & $-\smfrac{3}{16}$      & $-\smfrac{47}{64}$   
         & $\smfrac{101}{384}$     \\
$ (F_0 - \gamma_E)/4 \pi^2 $ & $-3$ & $-3$ & 
         $0$       & $0$       & $1$ \\          
$Y_0$ 
         & $- 12$ & $- 12$ & $- 2$ & $- 2$                
         & $\smfrac{4}{3}$  \\
$Y_1$
         & $- 18$ & $-6$    
         & $\smfrac{1}{2}$       &  $-\smfrac{11}{2}$      
         & $\smfrac{2}{3}$     \\
$Y_2$
         & $- \smfrac{1}{4}$ & $- \smfrac{1}{4}$ 
         &  $ \smfrac{3}{8}$ & $ \smfrac{3}{8}$ 
         & $ \smfrac{7}{12}$ \\
$Y_3$
        & $- 48$              &$- 48$              
        & $- 48$              &$- 48$              
        & $- 48$              \\
$Z_0$
        & $6$              & $2$              
        & $- 1$            & $1$              
        & $-\smfrac{4}{3}$             \\      
$Z_1$
        & $- \smfrac{3}{4}$ & $- \smfrac{3}{4}$            
        & $- \smfrac{3}{4}$ & $- \smfrac{3}{4}$            
        & $- \smfrac{3}{4}$             \\      
${\cal F}(1,-2)$
        & $\smfrac{4}{3}$ & $\smfrac{7}{96}$
        & $-\smfrac{11}{48}$ & $\smfrac{77}{192}$ 
        & $-\smfrac{23}{192}$ \\
${\cal F}(1,-1)$
        & $-\smfrac{329}{192}$ & $-\smfrac{139}{192}$
        & $\smfrac{77}{384}$ & $-\smfrac{113}{384}$
        & $\smfrac{11}{32}$ \\
${\cal F}(1, 0)$
        & $\smfrac{157}{8}$ & $-\smfrac{99}{8}$ 
        & $-\smfrac{1373}{96}$ & $\smfrac{163}{96}$
        & $-\smfrac{173}{18}$ \\
${\cal F}(2,-2)$
        & $\smfrac{4321}{192}$ & $-\smfrac{271}{192}$
        & $-\smfrac{2737}{384}$ & $\smfrac{1855}{384}$
        & $-\smfrac{323}{64}$ \\
${\cal F}(2,-1)$
        & $-\smfrac{1823}{8}$ & $\smfrac{117}{8}$
        & $\smfrac{1693}{24}$ & $-\smfrac{1217}{24}$
        & $\smfrac{3511}{72}$   \\
${\cal F}(3,-4)$
        & $-\smfrac{7}{32}$ & $0$
        & $\smfrac{7}{128}$ & $-\smfrac{7}{128}$
        & $\smfrac{7}{192}$ \\
${\cal F}(3,-3)$
        & $-\smfrac{1585}{32}$ & $\smfrac{105}{32}$
        & $\smfrac{1055}{64}$ & $-\smfrac{635}{64}$
        & $\smfrac{145}{12}$ \\
${\cal F}(3,-2)$
        & $\smfrac{4713}{16}$ & $-\smfrac{203}{16}$
        & $-\smfrac{179}{2}$ & $\smfrac{513}{8}$
        & $-\smfrac{3067}{48}$  \\
Total
        & $-0.2634473$ & $-0.5669576$
        & $-0.2493438$ & $-0.0975887$
        & $-0.0591138$  \\
\hline
Ref. \protect\cite{bilinearfermions} 
        & $-0.269$ & $-0.573$ & $-0.249$ & $-0.0973$ & $-0.0570$ \\
\hline
\end{tabular}
\end{center}
\caption{Results for $\Delta Z_\Gamma$ for the improved 
operators \protect\reff{eq5.11}: we report here the coefficients 
of the various constant appearing in the result. ``Total" is 
the numerical value of $\Delta Z_\Gamma$.}
\label{Z_bilineari_improved}
\end{table}

\renewcommand{\arraystretch}{1}

\subsubsection{Renormalization constants for fermionic energy-momentum
tensor}

We want now to compute the renormalization constants for the dimension-four
operators which appear in the first moment of the 
deep-inelastic-scattering structure functions  and in the definition
of the energy-momentum tensor. We will consider two operators 
whose continuum {\em formal} limit is 
\begin{eqnarray}
{\cal O}^{(1)}_{\mu \nu} &=& {1\over 4} \left[
   \overline\psi \gamma_\mu D_\nu \psi - D_\nu\overline\psi \gamma_\mu \psi +
    \, \, (\mu\leftrightarrow\nu) \right] \\
{\cal O}^{(2)}_{\mu \nu} &=& 
    F^a_{\mu\alpha} F^a_{\nu\alpha} 
\label{FmaFna}
\end{eqnarray}
For their explicit definition in terms of lattice operators 
we refer to \cite{CCMP-Annals}, formula (5.22) and 
\cite{CMP}, formulae (3.14)/(3.16). We want now to compute new operators 
so that 
\be
   \< f| {\widehat{\cal O}}^{(i),LATT}_{\mu\nu} |i\> \, =\, 
   \< f| {\cal O}^{(i),CONT}_{\mu\nu} |i\>
\ee
for arbitrary states $f$ and $i$. In the continuum we adopt 
the $\overline{MS}$-scheme with scale $\mu=1/a$. To define 
${\widehat{\cal O}}^{(i),LATT}_{\mu\nu}$ we must consider all possible 
mixings with operators of dimension less than or equal to four. Here we will
restrict our attention to the gluonic sector. We will write at one loop
\be
   {\widehat{\cal O}}^{(i),LATT}_{\mu\nu} = \, 
    {\cal O}^{(i)}_{\mu\nu} + g^2 \Delta^{(i)}_{\mu\nu}
\ee
where 
\begin{eqnarray}
\Delta^{(i)}_{\mu\nu} &=& \Delta^{(i)}_1 
      \left( F^a_{\mu\alpha} F^a_{\nu\alpha} - 
       {1\over 4} F^2 \delta_{\mu\nu}\right)^{LATT} + 
       \Delta^{(i)}_2 \left( {1\over 4} F^2 \delta_{\mu\nu}\right)^{LATT} 
\nonumber \\
      && + \Delta^{(i)}_3 \delta_{\mu\nu}
      \left( F^a_{\mu\alpha} F^a_{\nu\alpha} -
       {1\over 4} F^2 \delta_{\mu\nu}\right)^{LATT}
\end{eqnarray}
The superscript $LATT$ indicates that we use here some lattice operator 
with the given continuum limit: the explicit discretization is however
irrelevant at one loop. For $r_W=1$ the constants $\Delta^{(1)}_i$ 
are reported in table \ref{emt_table}. For the gluonic operator 
\reff{FmaFna} we write $\Delta^{(2)}_i = \Delta^{(2g)}_i + N_f 
\Delta^{(2f)}_i$. The constants $\Delta^{(2g)}_i$ have been computed in 
\cite{CMP}, see formula (6.11)\footnote{The calculation in 
\cite{CMP} considers minimal
subtraction in the continuum. If one considers the 
$\overline{MS}$-scheme, one should use the formulae of \cite{CMP},
sect. 6, with $Y=(F_0-\gamma_E)/\pi^2$.}. As for the constants 
$\Delta^{(2f)}_i$, we have $\Delta^{(2f)}_3=0$, $\Delta^{(2f)}_2 = 
\Delta^{(2f)}_1$. The explicit expression of 
$\Delta^{(2f)}_1$ for $r_W=1$ is given  in table \ref{emt_table}.

\renewcommand{\arraystretch}{1.5}

\begin{table}
\begin{center}
\begin{tabular}{|l|rrrr|}
\hline
 & \multicolumn{1}{r}{$\Delta^{(1)}_1$}   
 & \multicolumn{1}{r}{$\Delta^{(1)}_2$}     
 & \multicolumn{1}{r}{$\Delta^{(1)}_3$}     
 & \multicolumn{1}{r|}{$\Delta^{(2f)}_1$}     \\
\hline
$1$ & $\smfrac{2}{27}$ & $-\smfrac{1}{108}$ 
         & $-\smfrac{11}{108}$    &    $\smfrac{1}{9}$            \\
$1/\pi^2$ & $-\smfrac{11}{144}$ & $-\smfrac{1}{288}$  
         & $\smfrac{19}{96}$      & $0$    \\
$ (F_0 - \gamma_E)/4\pi^2 $ & $-\smfrac{1}{3}$ & $0$ & 
         $0$       & $\smfrac{1}{3}$      \\ 
$Y_0$ 
         & $-\smfrac{4}{3}$ & $0$ & $0$ & $\smfrac{4}{3}$     \\
${\cal F}(1,-2)$
        & $-\smfrac{1}{54}$ & $\smfrac{1}{432}$
        & $\smfrac{11}{432}$ & $-\smfrac{1}{36}$ \\
${\cal F}(1,-1)$
        & $-\smfrac{2}{27}$ & $\smfrac{25}{432}$
        & $\smfrac{47}{432}$ & $-\smfrac{1}{18}$  \\
${\cal F}(1, 0)$
        & $\smfrac{13}{9}$ & $\smfrac{7}{36}$ 
        & $-\smfrac{137}{36}$ & $\smfrac{7}{6}$  \\
${\cal F}(2,-2)$
        & $-\smfrac{1}{54}$ & $-\smfrac{97}{216}$
        & $\smfrac{43}{216}$ & $-\smfrac{5}{24}$  \\
${\cal F}(2,-1)$
        & $-\smfrac{62}{9}$ & $\smfrac{13}{6}$
        & $\smfrac{359}{18}$ & $-\smfrac{2}{3}$ \\
${\cal F}(3,-4)$
        & $0$ & $\smfrac{13}{288}$
        & $-\smfrac{13}{288}$ & $0$  \\
${\cal F}(3,-3)$
        & $0$ & $\smfrac{1}{8}$
        & $-\smfrac{17}{72}$ & $0$  \\
${\cal F}(3,-2)$
        & $\smfrac{149}{18}$ & $-\smfrac{113}{36}$
        & $-\smfrac{791}{36}$ & $0$  \\
Total
        & $0.00826199$ & $-0.01058036$
        & $-0.00963232$ & $0.01339200$    \\
\hline
\end{tabular}
\end{center}
\caption{Results for $\Delta^{(i)}_j$: we report here the coefficients 
of the various constant appearing in the result. ``Total" is 
the numerical value. All constants must be multiplied by
$T_f$ defined by Tr$\ T^a T^b = T_f \delta^{ab}$. For fermions in the
fundamental representation of $SU(N)$ we have $T_f = \smfrac{1}{2}$.}
\label{emt_table}
\end{table}
\renewcommand{\arraystretch}{1}

We can also easily compute the renormalization constants for the 
operators which are the trace of 
${\cal O}^{(1)}_{\mu\nu}$ and ${\cal O}^{(2)}_{\mu\nu}$, i.e. for
\begin{eqnarray}
{\cal O}^{(3)} &=& {1\over 2} \sum_\mu \left( \overline \psi 
    \gamma_\mu D_\mu \psi - D_\mu \overline\psi \gamma_\mu \psi \right)
\\
{\cal O}^{(4)} &=& F^a_{\alpha\beta} F^a_{\alpha\beta}
\end{eqnarray}
Considering again only the gluonic sector\footnote{
As before there are additional mixings with dimension-three and dimension-four
fermionic operators. For a numerical evaluation of these mixings see 
\cite{Capitani-Rossi,Desy}.} we can write 
\begin{eqnarray}
   \widehat{{\cal O}}^{(3)} &=& {\cal O}^{(3)} + {g^2} 
     \Delta^{(3)} {\cal O}^{(4)} \\
    \widehat{{\cal O}}^{(4)} &=&
   \left( 1 + {g^2} \Delta^{(4)}\right) {\cal O}^{(4)}
\end{eqnarray}
We have\footnote{The additional terms which appear in 
$\Delta^{(3)}$ and $\Delta^{(4)}$ are due to the fact that in 
dimensional regularization $\delta_{\mu\nu} N({\cal O}_{\mu\nu}) \not=
N(\delta_{\mu\nu} {\cal O}_{\mu\nu})$ \cite{Breitenlohner,Bonneau}.} 
\begin{eqnarray}
    \Delta^{(3)} &=&  \Delta^{(1)}_2 + {1\over 24 \pi^2} T_f\\
    \Delta^{(4)} &=&  \Delta^{(2g)}_2 + \Delta^{(2f)}_2 - 
          {11 N\over 96 \pi^2}
\end{eqnarray}
where $T_f$ is defined by Tr$\ T^a T^b = T_f \delta^{ab}$ 
($T_f=\, \smfrac{1}{2}$ for fermions in the fundamental representation of 
$SU(N)$).
The last numbers in $\Delta^{(3)}$ and $\Delta^{(4)}$ are related to the 
anomaly of the energy-momentum tensor. Indeed define
\be
T_{\mu\nu} \, =\, \sum_f \left(  \widehat{{\cal O}}^{(1),f}_{\mu\nu} 
     - {1\over4} \delta_{\mu\nu} \widehat{{\cal O}}^{(3),f} \right) +\, 
    \widehat{{\cal O}}^{(2)}_{\mu\nu} - 
    {1\over4} \delta_{\mu\nu} \widehat{{\cal O}}^{(4)}
\ee
where the first sum is over the $N_f$ fermion species. Then 
\be
     T_{\mu\mu} = g^2\left( {11 N\over 96 \pi^2} - 
         {1\over 24 \pi^2} N_f T_f \right) F^2
\ee
in agreement with \cite{anomalia1,anomalia2}.

Finally we want to compare our results with those of Capitani and Rossi
\cite{Capitani-Rossi}.
An easy calculation gives:
\begin{eqnarray}
      B_{gg} &=& - 16 \pi^2 {\Delta^{(2g)}_1\over N} - 
                 {4 \over 3} \approx  - 17.778285 + {2\pi^2\over N^2} \\
      B^f_{gg} &=& - 16 \pi^2 \Delta^{(2f)}_1 - {20\over 9}T_f 
               \approx - 2.16850094 \, (2 T_f) \\
      B_{qg} &=& 16 \pi^2 \Delta^{(1)}_1 - {8\over 9} T_f \approx 
              0.20789614 \, (2 T_f)
\end{eqnarray}

As already noticed in \cite{Capitani-Rossi} our final result for $B_{gg}$
agrees with theirs\footnote{There are however some misprints in Tables 
10 and 11 of \cite{Capitani-Rossi}: in Table 10 the contribution of 
``{Sails}" is 
$-85/(144 \pi^2) - 7/(24 \pi^2) (2/\epsilon + \log 4\pi - \gamma_E)$ and 
``Total $I-J$" is ($-1/(12 \pi^2)$), while in Table 11 in 
``{Sails}" $-7/(9 \pi^2)$ should be replaced by $-3/(16 \pi^2)$ and 
in ``Total $J$", $-11/(18 \pi^2)$ should be $-1/(48 \pi^2)$.
We are also in disagreement with their Table 2, where we get 
$B_{gg} = - 4/3$, $B^f_{gg} = - 20 T_f/9$, $B_{qg} = - 8 T_f/9$.
However a recomputation confirms our results~\cite{CR-private}.}. 
We are also in perfect agreement for $B^f_{gg}$, while we differ for 
$B_{qg}$ which is reported in \cite{Capitani-Rossi} to be 
$B_{qg} = 0.019$ for $T_f = \smfrac{1}{2}$\footnote{
This discrepancy has been recently understood and the new result is
in agreement with ours~\cite{CR-private}.}.

We want also to correct here the results which appeared in \cite{CCMP-PLB}.
Indeed the constants $Z_5$, $Z_6$ and $Z_7$ appearing in Table 1 should be
$Z_5 = 1 + g^2 1.02165$, $Z_6 = - g^2 0.65205$ 
and $Z_7 = g^2 0.25034$.
The different result for $Z_5$ and $Z_7$ was due to a numerical mistake
in the evaluation of the photon contribution (the exact result is 
in \cite{CMP}, sec. 5). There are also minor misprints: 
the sign on $F^2$ in (16), (17) and (18) of \cite{CCMP-PLB} should be
``minus". An analogous sign should be changed in section 7 of 
\cite{CCMP-Annals}.
      
\subsection{Calculation of the propagators}

Recently a very efficient numerical method to evaluate higher-loop
integrals has been presented in \cite{LW}. An essential ingredient 
in the method is the exact calculation of the free bosonic 
propagator. In \cite{LW} an algorithm was introduced which allowed to 
express its values in $x$-space in terms of the values at 
$x=(0,0,0,0)$ and $x=(1,1,0,0)$. Here we will discuss how to use our
method to obtain an algorithm for the bosonic and the Wilson-fermion
propagator.

In general we will consider
\begin{eqnarray}
G_B (p;x) = \int {dk\over (2 \pi)^4} {e^{ikx}\over D_B(k,m)^p} 
\label{GB}\\
G_F (p,q;x) =\, \int {dk\over (2\pi)^4} 
{e^{ikx}\over \hat{D}_F(k,m)^p D_B(k,m)^q}
\label{GF}
\end{eqnarray}
The computation of these two quantities is in principle straightforward:
indeed, using the symmetry $k_\mu \to -k_\mu$ we can rewrite 
$e^{ikx}$ as $\prod_\mu \cos (k_\mu x_\mu)$ and then express 
$\cos (k_\mu x_\mu)$ as a polynomial in $\hat{k}^2_\mu$. In this way
$G_B(p;x)$ and $G_F(p,q;x)$ are expressed\footnote{Notice that also the 
reverse is true: the integrals ${\cal B}$ and ${\cal F}$ 
can be rewritten in terms of $G_B$ and $G_F$ using the identity 
$(\hat{k}^2)^m = (-1)^m (e^{ik/2} - e^{-ik/2})^{2m} $. Loosely 
speaking $G_B$ and ${\cal B}$ (and analogously $G_F$ and ${\cal F}$) 
are two different ``basis" in the space
of bosonic (respectively mixed bosonic-fermionic) lattice integrals.}
in terms of ${\cal B}$ and 
$\hat{\cal F}$ and thus, for $m\to 0$, using the results of the previous
Sections, we can express $G_B(p;x)$ in terms of $Z_0$ and $Z_1$ 
(and $F_0$ if $p\ge 2$) and $G_F(p,q;x)$ in terms of the eight constants
${\cal F}(1,[-2,-1,0])$, ${\cal F}(2,[-1,-2])$ and 
${\cal F}(3,[-4,-3,-2])$ if $q\le 0$ and $p=1$, to which we must add 
${\cal F} (2,0)$ if $q\le 0$, $p\ge 2$, and $Y_0$, $Y_1$, $Y_2$, $Y_3$,
$Z_0$ and $Z_1$ if $q>0$.
Of course, in a completely equivalent way, in the bosonic
case we can use instead of $Z_0$ and $Z_1$ the values of 
$G_B(1;x)$ at two different values of $x$. For instance, as 
in \cite{LW}, we could choose
\begin{eqnarray}
G_B(1;(0,0,0,0)) &=& Z_0 \\
G_B(1;(1,1,0,0)) &=& Z_0 + Z_1 - {1\over4}
\end{eqnarray}
Analogously in the fermionic case we could express $G_F(p,q;x)$ for 
$q\le 0$ in terms
of the values of $G_F(1,0;x)$ at eight different points: 
a possible choice, with all the points in a hypercube of side two,
is given by $(0,0,0,0)$, $(1,0,0,0)$, $(2,0,0,0)$,
$(2,1,0,0)$, $(2,2,0,0)$, $(2,2,1,0)$, $(2,2,2,0)$ and $(1,1,1,1)$.

{}From a practical point of view, one can simplify the 
algorithm by implementing the ``integration-by-part"
recursion relations \reff{rec2} and \reff{fermionicrec3} 
directly on the propagators. To apply our technique
we will generalize \reff{GB} and \reff{GF} by considering
\begin{eqnarray}
{\cal G}_B (p;x) &=& \int {dk\over (2\pi)^4} {e^{ikx}\over D_B(k,m)^{p+\delta}}
    \\
{\cal G}_F (p,q;x) &=& \int {dk\over (2\pi)^4}
	{e^{ikx}\over \hat{D}_F(k,m)^{p+\delta} D_B(k,m)^q}
\end{eqnarray}
For these two quantities it is very easy 
to obtain recursion relations. We obtain in the two cases:
\begin{eqnarray}
{\cal G}_B(p;x+\mu) &=& {\cal G}_B(p;x-\mu) - {x_\mu \over p+\delta-1}
       {\cal G}_B (p-1;x) \label{bosonicproprecursion}\\
{\cal G}_F(p,q;x+2\mu) &=& {\cal G}_F (p,q;x-2\mu) -
   r^2_W \left( {\cal G}_F(p,q-1;x+\mu) - {\cal G}_F(p,q-1;x-\mu)\right) 
   \nonumber \\ [1mm]
   && + (m^2 r_W^2- 2m) \left( {\cal G}_F(p,q;x+\mu) -
	{\cal G}_F (p,q;x-\mu)\right) \nonumber \\
   && - {2\over p - 1 +\delta} \left[x_\mu
	 {\cal G}_F (p-1,q;x) \right. \nonumber \\
   && \quad + q \left. \left(
	{\cal G}_F (p-1,q+1;x+\mu) - {\cal G}_F (p-1,q+1;x-\mu)\right)\right]
\end{eqnarray}
where $\mu$ is a lattice unit vector.
Using these two relations it is possible to express any element
${\cal G}_B (p;x)$ in terms of ${\cal G}_B (p';x')$ where $x'$ is an
element of the unit hypercube (i.e. ${x'}_\mu = 0$ or 1).
Analogously in the fermionic case we express  every
${\cal G}_F(p,q;x)$ in terms of ${\cal G}_F(p',q';x')$
with $x'$ belonging to the
hypercube of side two, i.e. ${x'}_\mu = 0$, 1 and 2. These last quantities
can then be easily expressed in terms of ${\BB}$ or ${\FF}$ and then
the procedure we have been presented in the previous Sections can be
applied. 
In the bosonic case, the algorithm we have described is less efficient
than the one introduced by \cite{Vohwinkel}: indeed for $p=1$,
starting from \reff{bosonicproprecursion}, Vohwinkel obtains a 
recursion relation which involves only terms with $p=1$ and which 
thus avoids the necessity of introducing $\delta$ and nonpositive
values of $p$.
In the fermionic case however we have not been able to implement the same 
trick.

Finally let us notice that once \reff{GF} has been computed one can easily 
obtain the fermion propagator
\be
\Delta_F(x) \, =\, 
   \int {d^4 k\over (2\pi)^4} e^{ikx}\, 
    {- i (\sum_\mu \gamma_\mu \sin k_\mu) + \smfrac{1}{2} r_W \hat{k}^2 + m 
    \over \hat{D}_F (k,m) }
\ee
as 
\begin{eqnarray}
\Delta_F(x)  &=& - {1\over2} \sum_\mu \gamma^\mu
 \left( G_F(1,0;x+\mu) - G_F(1,0;x-\mu) \right) - 
\nonumber \\ 
&& \qquad {r_W\over2} 
 \sum_\mu \left(G_F(1,0;x+\mu) + G_F(1,0;x-\mu) \right) \,
+ (4 r_W + m) G_F(1,0;x) \nonumber \\ [-2mm]
{}
\end{eqnarray}

To conclude the section let us discuss the numerical evaluation of the 
fermionic propagator. As in the bosonic case \cite{LW}, the expressions
for $G_F(p,q;x)$ in terms of our basic constants are numerically unstable
for $|x|\to\infty$: a numerical error in the basic constants gets 
amplified in $G_F(p,q;x)$ as $|x|\to\infty$. This problem has a 
standard way  out: if the expressions are unstable going outward from 
the origin, they will be stable in the opposite direction: thus,
if we want to compute the propagator for $|x|<d$, for some fixed $d$, we 
choose eight points with $|x|\approx d$ (say $y_1,\ldots,y_8$) and 
then we express the propagator for $|x|<d$ in terms of 
$G_F(1,0;y_i)$, $i=1,\ldots,8$. The new expressions are numerically stable:
the numerical error on $G_F(1,0;y_i)$ gets 
{\em reduced} when we compute the propagator
for $|x|\to 0$. As noticed in \cite{LW} the instability of the recursion
can also be used to provide precise estimates for the basic constants.
We have thus used this method to obtain an independent numerical estimate 
of the first eight constants of table \ref{costanti_fermioniche} 
considering the set of eight points 
$X^{(n)}\equiv \{(n,[0-3],0,0),(n+1,[0-3],0,0)\}$.
If one estimates the propagator $G_F(1,0;x)$ 
at $X^{(n)}$ with an accuracy of 
$1\%$, one reproduces the results of Table 
\ref{costanti_fermioniche} with an accuracy of 
approximately $10^{-6}$ (respectively $10^{-8}$) with 
$n=7$ ($n=9$ resp.).

It is also possible to apply the same procedure to $G_F(1,q;x)$ with 
$q<0$. The main advantage is that, using larger
negative values of $q$, one can obtain more precise estimates of 
$G_F(1,q;x)$ at the set of points $X^{(n)}$. With $q=-2$ and $n=7$,
using the method presented at the end of section \ref{sec3.4}
(i.e. evaluating the integrals by computing discrete sums of the 
form \reff{estrapolazionenumerica} with $L=50$ -- $100$), we obtain 
the values of $G_F(1,-2;x)$ at the points $X^{(n)}$ with a relative precision
of $10^{-6}$ and the final estimates of the basic constants with a 
precision of $10^{-9}$. If one needs better estimates one can obtain them
by simply increasing both $n$ and $-q$. The precision can in this way 
be increased at will. Using $n=26$ and $q=-3$ we checked the first
eight constants of table \ref{costanti_fermioniche}.

\section*{Acknowledgments}
We want to thank Pietro Menotti for collaborating in the early
stages of this work. We also thank Stefano Capitani and Gian Carlo Rossi for 
useful correspondence.

\appendix

\section{Final relations}

In this appendix we report the relations which have been found in the fourth
step of our reduction procedure for the fermionic integrals.
The relations for ${\FF}(p,q)$, $q\le 0$ are
\begin{eqnarray}
&& \hskip -40pt 
   \FF(0, -4) = 7336 \, \FF(0,0) +\delta\left[ -{910433 \over 144 \pi^2}  - 
   {1024087\over36}  + {325123 \over 144} \FF(1, -2)
   \right. \nonumber \\ 
   && \quad + 
   {2538989 \over72} \FF(1, -1) + {501607\over6} \FF(1, 0) - 
   {14398667 \over36} \FF(2, -2) + {3075865 \over9} \FF(2, -1) 
    \nonumber \\ 
    && \quad \left. + 
   {1919221\over48}  \FF(3, -4) + {7343317\over12} \FF(3, -3) - 
   {20889553\over18} \FF(3, -2) \right]
\\
&& \hskip -40pt
\FF(0, -3) = 704\, \FF(0,0) + \delta\left[ {9029 \over9 \pi^2} 
   - {12392 \over27}  - 
   {13138 \over 27} \FF(1, -2) 
   \right. \nonumber \\ 
   && \quad  + {115898 \over27} \FF(1, -1) - 
   {102632 \over3} \FF(1, 0) - {1579072\over27} \FF(2, -2) + 
   {826640 \over3} \FF(2, -1) 
    \nonumber \\ 
    && \left. \quad + 1190 \, \FF(3, -4) + 
   {369580 \over3} \FF(3, -3) - {3903064\over9} \FF(3, -2) \right ]
\\
&& \hskip -40pt
\FF(0, -2) = 72\, \FF(0,0) + \delta \left[ {35\over 4 \pi^2} - 
      {443\over3}  - 
   {7\over12} \FF(1, -2) + {1657\over6} \FF(1, -1) 
   \right. \nonumber \\ 
   && \quad  -\, 678\, \FF(1, 0) - 
   {8173\over3} \FF(2, -2) + 8340\, \FF(2, -1) + {525\over4} \FF(3, -4) 
    \nonumber \\ 
   && \quad \left. + 
   5265\, \FF(3, -3) - 15322\, \FF(3, -2) \right]
\\ [2mm]
&& \hskip -40pt
\FF(0, -1) = 8\, \FF(0,0) + \delta\left[ {15 \over 2 \pi^2} 
        - 4\, \FF(1, -2) + 
   25\, \FF(1, -1) - 228\, \FF(1, 0)
   \right. \nonumber \\ 
   && \quad \left.  - 268\, \FF(2, -2) + 
   1608\, \FF(2, -1) + 630\, \FF(3, -3) - 2436\, \FF(3, -2) \right]
\\ [2mm]
&& \hskip -40pt
\FF(1, -4) = -{1606\over3 \pi^2} - {6088\over9} + {2962\over9} \FF(1, -2) - 
   {3548\over9} \FF(1, -1) 
   \nonumber \\
&& \quad + 14992\, \FF(1, 0) + {100312\over9} \FF(2, -2) - 
   {282272\over3} \FF(2, -1) + {1366\over3} \FF(3, -4) 
   \nonumber \\ 
&& \quad - {86456\over3} \FF(3, -3) + 
   {388304\over3} \FF(3, -2)
\\
&& \hskip -40pt
\FF(1, -3) = {17\over\pi^2} + {208 \over3} - {52\over3} \FF(1, -2) + 
           {89\over3} \FF(1, -1) - 528 \, \FF(1, 0) 
   \nonumber \\
&& \quad 
- {1756\over3} \FF(2, -2) + 3696\, \FF(2, -1) + 1366\, \FF(3, -3) - 
   5368\, \FF(3, -2)
\\
&& \hskip -40pt
\FF(2, -4) = {13\over\pi^2} - {52\over3} + {31\over3} \FF(1, -2) 
    - {38\over3} \FF(1, -1) - 264\, \FF(1, 0) 
    \nonumber \\
&& \quad + {52\over3} \FF(2, -2) + 1200\, \FF(2, -1) - \FF(3, -4) + 4\, \FF(3, -3) - 
   1304\, \FF(3, -2)
\\
&& \hskip -40pt
\FF(2, -3) = -{3\over 2 \pi^2} + 5\, \FF(1, -1) + 36\, \FF(1, 0) - 4\, \FF(2, -2) 
\nonumber \\ 
&& \quad - 
   168\, \FF(2, -1) - 2\, \FF(3, -3) + 180\, \FF(3, -2)
\\ [2mm]
&& \hskip -40pt
\FF(3, -1) = -{31\over 1024 \pi^2} + {11\over128} \FF(1, 0) 
  - {83\over192} \FF(2, -1) + 
   \FF(2, 0) + {155\over384} \FF(3, -2)
\\
&& \hskip -40pt
\FF(3, 0) = {1\over 32 \pi^2 m^2} + {1433\over 49152 \pi^2}  
   - {1\over 3072} \FF(1, -1) - 
   {461\over 6144} \FF(1, 0) + {85\over768} \FF(2, -2) 
\nonumber \\
&& \quad + {1405\over9216} \FF(2, -1) - 
   {1\over4} \FF(2, 0) - {35\over2048} \FF(3, -4) - 
   {57\over512} \FF(3, -3) - {301\over18432} \FF(3, -2) \nonumber \\ [-2mm]
&& {}
\end{eqnarray}
For positive $q$ we have
\begin{eqnarray}
&& \hskip -40 pt
\FF(2,1) \,= \,
   {{427}\over {18432 \pi^2}} 
   + {1\over {48}} \FF(0,2)  
   -{1\over {384}} \FF(0,1)  
   + \FF(1,2) -{5\over {12}} \FF(1,1)  \nonumber \\
&&   -{{19}\over {768}} \FF(1,0)  
   -{1\over {9216}} \FF(1,-1) 
   + {{13}\over {48}} \FF(2,0)
   + {{35}\over {288}} \FF(2,-1) 
   + {{85}\over {2304}} \FF(2,-2) \nonumber \\
&&   -{{35}\over {6144}} \FF(3,-4)  
   -{{19}\over {512}} \FF(3,-3)  
   -{{173}\over {2304}} \FF(3,-2)  \\
&& \hskip -40 pt
\FF(2,2) \, = \, 
    {1\over {96\pi^2 m^4}}
   -{{15}\over {1024\pi^2 m^2}}
   -{{2677}\over {11059200\pi^2}}
   -{1\over {30}} \FF(0,3) 
   + {{41}\over {2880}} \FF(0,2) \nonumber \\
&&  -{{67}\over {46080}} \FF(0,1) 
   + {{221}\over {480}} \FF(1,2) 
  -{{2383}\over {11520}} \FF(1,1) 
   + {{403}\over {18432}} \FF(1,0)  \nonumber \\
&&   + {{23}\over {276480}} \FF(1,-1) 
   + {{2681}\over {11520}} \FF(2,0)  
   -{{4493}\over {276480}} \FF(2,-1)  
   -{{391}\over {13824}} \FF(2,-2)  \nonumber \\
&&   -{{4433}\over {276480}} \FF(3,-2)  
   + {{437}\over {15360}} \FF(3,-3)  
   + {{161}\over {36864}} \FF(3,-4)   \\
&& \hskip -40 pt
\FF(1,3) \, = \,
   {1\over {96\pi^2 m^4}}
   -{3\over {1024\pi^2 m^2}}
   -{{214253}\over {58982400\pi^2}}
  -{{13}\over {320}} \FF(0,3) 
   + {1\over {3840}} \FF(0,2) \nonumber \\
&&  + {{47}\over {92160}} \FF(0,1)  
   + {{149}\over {960}} \FF(1,2)  
   + {{391}\over {7680}} \FF(1,1)  
   + {{169}\over {294912}} \FF(1,0)  \nonumber \\
&& + {7\over {245760}} \FF(1,-1) 
   -{{37}\over {1280}} \FF(2,0)  
   + {{323}\over {245760}} \FF(2,-1)  
   -{{119}\over {12288}} \FF(2,-2)  \nonumber \\
&& -{{6139}\over {491520}} \FF(3,-2)  
   + {{399}\over {40960}} \FF(3,-3)  
   + {{49}\over {32768}} \FF(3,-4)  \\
&& \hskip -40 pt
\FF(1,4) \, = \,
    {1\over {192\pi^2 m^6}}
   + {1\over {3840\pi^2 m^4}}
   + {{43}\over {69120\pi^2 m^2}}
   -{{230795603}\over {277453209600\pi^2}} \nonumber \\
&& -{{22817}\over {1935360}} \FF(0,3)   
   -{{3529}\over {3870720}} \FF(0,2) 
   + {{509}\over {1720320}} \FF(0,1) 
   -{{533}\over {645120}} \FF(1,2) \nonumber \\
&& + {{164753}\over {5160960}} \FF(1,1) 
   -{{2823307}\over {990904320}} \FF(1,0) 
   -{{19}\over {55050240}} \FF(1,-1)  
  -{{25471}\over {860160}} \FF(2,0) \nonumber \\
&& + {{8917219}\over {1486356480}} \FF(2,-1)  
   + {{323}\over {2752512}} \FF(2,-2)  
   -{{17909419}\over {2972712960}} \FF(3,-2)  \nonumber \\
&& -{{1083}\over {9175040}} \FF(3,-3)  
   -{{19}\over {1048576}} \FF(3,-4) \\
&& \hskip -40 pt
\FF(0,4) \, = \, (1 - \delta \log m^2) \left[
   {1\over {96\pi^2 m^4}}
   -{{19}\over {4608\pi^2 m^2}}
   + {1\over {9216\pi^2}}\right] \nonumber \\ 
&& + {{31}\over {144}} \FF(0,3)  
  -{{13}\over {1152}} \FF(0,2) 
   + {1\over {9216}} \FF(0,1) \nonumber \\
&& \hskip -10pt
+ \delta \left[
   -{5\over {576\pi^2 m^4}}
   + {{61}\over {18432\pi^2 m^2}}
   + {{80989}\over {88473600\pi^2}}
   + {{347}\over {1440}} \FF(0,3) \right. \nonumber \\
&&   -{{83}\over {2560}} \FF(0,2)  
   + {{137}\over {184320}} \FF(0,1) 
  -{{689}\over {2880}} \FF(1,2) 
   + {{1139}\over {23040}} \FF(1,1) \nonumber \\
&&   -{{415}\over {147456}} \FF(1,0) 
   + {{23}\over {4423680}} \FF(1,-1)  
   -{{329}\over {11520}} \FF(2,0)  
   + {{13283}\over {1105920}} \FF(2,-1) \nonumber \\
&& \left.   
   -{{391}\over {221184}} \FF(2,-2)  
   -{{30479}\over {2211840}} \FF(3,-2)  
   + {{437}\over {245760}} \FF(3,-3) 
   + {{161}\over {589824}} \FF(3,-4)  
  \right] \nonumber \\ [-2mm]
&& {} \\
&& \hskip -40 pt
\FF(0,5) \, = \, (1 - \delta \log m^2) \left[
   {1\over {192\pi^2 m^6}} 
   + {1\over {1536\pi^2 m^4}} 
   -{{415}\over {442368\pi^2 m^2}} 
   + {{11}\over {442368\pi^2}} \right] \nonumber \\
&&   + {{523}\over {13824}} \FF(0,3) 
  -{{289}\over {110592}} \FF(0,2)  
   + {{25}\over {884736}} \FF(0,1)  \nonumber \\
&& \hskip -10pt
+ \delta \left[
   -{7\over {2304\pi^2 m^6}}
   -{{101}\over {92160\pi^2 m^4}}
  + {{1207}\over {1769472\pi^2 m^2}}
  + {{6492275537}\over {9988315545600\pi^2}} \right. \nonumber \\
&& + {{634603}\over {7741440}} \FF(0,3) 
   -{{2623}\over {258048}} \FF(0,2) 
  + {{61477}\over {371589120}} \FF(0,1) 
  -{{172157}\over {2580480}} \FF(1,2) \nonumber \\
&&  + {{515491}\over {61931520}} \FF(1,1) 
   -{{778597}\over {1321205760}} \FF(1,0)  
   -{{7297}\over {3567255552}} \FF(1,-1)  \nonumber \\
&& -{{57917}\over {10321920}} \FF(2,0) 
  + {{32033909}\over {17836277760}} \FF(2,-1)  
  + {{620245}\over {891813888}} \FF(2,-2) \nonumber \\ 
&& \left.   -{{7053481}\over {7134511104}} \FF(3,-2) 
   -{{138643}\over {198180864}} \FF(3,-3) 
   -{{36485}\over {339738624}} \FF(3,-4) 
 \right] \\
&& \hskip -40 pt
\FF(0,6) \, = \, (1 - \delta \log m^2) \left[
    {1\over {320\pi^2 m^8}} 
   + {1\over {3840\pi^2 m^6}}
   + {1\over {20480\pi^2 m^4}}
   -{{6209}\over {35389440\pi^2 m^2}} \right. \nonumber \\
&& \left.   + {{1091}\over {283115520\pi^2}}  \right]
   + {{1429}\over {221184}} \FF(0,3) 
  -{{4373}\over {8847360}} \FF(0,2)  
   + {{401}\over {70778880}} \FF(0,1)  \nonumber \\
&& \hskip -10pt
+ \delta \left[
   -{9\over {6400\pi^2 m^8}}
   -{{77}\over {230400\pi^2 m^6}}
   -{{709}\over {8601600\pi^2 m^4}}
   + {{24257}\over {1189085184\pi^2 m^2}} \right. \nonumber \\
&&   + {{903503350441}\over {4794391461888000\pi^2}}
   + {{74410051}\over {3715891200}} \FF(0,3) 
   -{{1058653}\over {464486400}} \FF(0,2)  \nonumber \\
&& + {{982249}\over {59454259200}} \FF(0,1) 
  -{{16644221}\over {1238630400}} \FF(1,2) 
   -{{2348599}\over {9909043200}} \FF(1,1) \nonumber \\
&&   + {{791158043}\over {5707608883200}} \FF(1,0) 
   -{{3838603}\over {8561413324800}} \FF(1,-1)  
   + {{3179779}\over {4954521600}} \FF(2,0) \nonumber \\
&&   -{{466704533}\over {1223059046400}} \FF(2,-1) 
   + {{65256251}\over {428070666240}} \FF(2,-2)  
   + {{9795427451}\over {17122826649600}} \FF(3,-2) \nonumber \\
&&\left.   -{{72933457}\over {475634073600}} \FF(3,-3) 
   -{{3838603}\over {163074539520}} \FF(3,-4)  
\right]
\end{eqnarray}

\end{document}